\documentclass{aa}
\usepackage{psfig,latexsym,longtable,lscape}

\def\>{$>$}
\def\<{$<$}

\def\newline{\hfil\break}
\def\mincir{\ \raise -2.truept\hbox{\rlap{\hbox{$\sim$}}\raise5.truept  
\hbox{$<$}\ }}                        
\def\magcir{\ \raise -2.truept\hbox{\rlap{\hbox{$\sim$}}\raise5.truept  
\hbox{$>$}\ }}                        
 
\def\bsax{BeppoSAX\,}

\def\e20{$\times 10^{20}$}
\begin{document}
 
   \thesaurus{3; 
              ( 11.09.1 M31; 
               11.19.2; 
               13.25.2) 
            }

\pagenumbering{arabic}
\title{Broad Band X--ray Spectra of M31 Sources with \bsax}

\author{Ginevra\,Trinchieri$^1$, Gian Luca Israel$^2$, Lucio Chiappetti$^3$,
Tomaso Belloni$^{1,4}$, 
Luigi Stella$^2$, Frank Primini$^5$, Pepi Fabbiano $^5$, Wolfgang Pietsch$^6$}

\institute{
Osservatorio Astronomico di Brera, via Brera 28, 20121
 Milano Italy
\and 
Osservatorio Astronomico di Roma, via Frascati 33, 00044 Roma Italy 
\and
Istituto di Fisica Cosmica ``G. Occhialini" (CNR), via Bassini 15,
Milano Italy
\and
Astronomical Institute ``A. Pannekoek" and Center for High Energy
Astrophysics, Kruislaan 403, 
1098 SJ Amsterdam,
The Netherlands
\and
Harvard-Smithsonian Center for Astrophysics, 60 Garden Street, Cambridge,
MA, 02138 USA
\and
Max-Planck-Institut f\"ur extraterrestrische Physik,
Giessenbachstrasse,
85748 Garching, Germany
}

   \mail{ginevra@brera.mi.astro.it}
   \offprints{G.~Trinchieri}

   \date{Received date; accepted date}
   \titlerunning{\bsax\ spectra of M31 sources  }
   \authorrunning{Trinchieri et al.}
   \maketitle
 
   \begin{abstract}

We present the first spectral study of the X--ray emitting 
stellar sources in M31 in the energy band from $\sim$ 0.1 to 10 keV.  
We find that the globular cluster sources have spectral
characteristics consistent with those of the Milky Way object, namely
that the spectrum can be described by a thermal model with $\sim 6-20
$ keV from $\sim$ 2 to 10 keV.  Evidence of high
absorption in some of these sources is most likely an indication
that they lie in or behind the HI ring in the disk of the galaxy.  
We also find one peculiar globular cluster source, with spectral
characteristics more typically associated with either High Mass X--ray
Binaries or black hole candidates.  We therefore suggest that either the
source has been wrongly identified with a globular cluster or
that the system contains a Black Hole.  

We confirm earlier report that the spectrum of the bulge of M31 is
consistent with the superposition of many LMXB spectra.  It is likely
that a large fraction of the $\sim 15-30$ keV detection obtained from
the PDS instrument is associated with the bulge, thus extending the
spectral data for this complex of sources up to $\sim 30$ keV.  The
high energy part of the spectrum can be parameterized with typical LMXB
spectra, while at low energies an additional component 
is required.

No significant variability is observed within the \bsax\ observation, 
while a few sources appear to have varied (brightened)
since ROSAT and $Einstein$ observations.

\keywords{Extragalactic astronomy; Galaxies: individual: M31; Galaxies:
spiral; X-rays: galaxies}

\end{abstract}

\section{Introduction}

\begin{figure*}

\resizebox{18cm}{!}{
\psfig{file=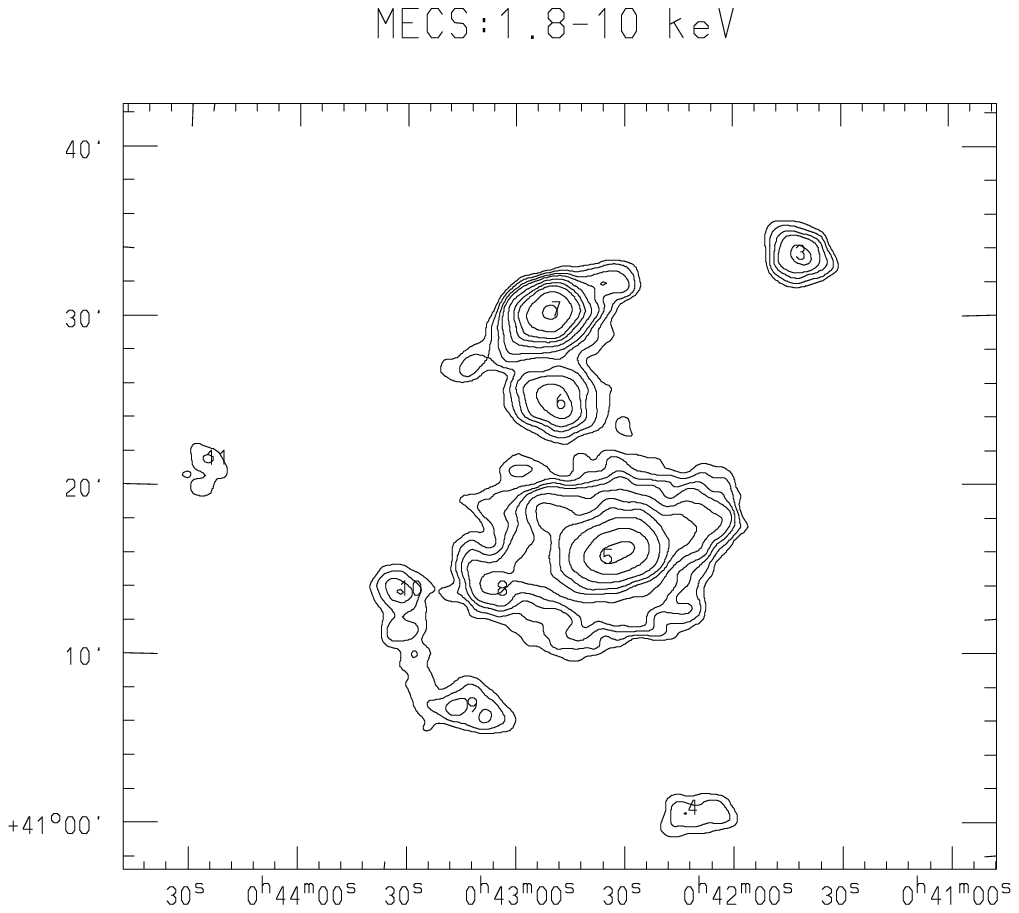,width=18cm,clip=}
\psfig{file= 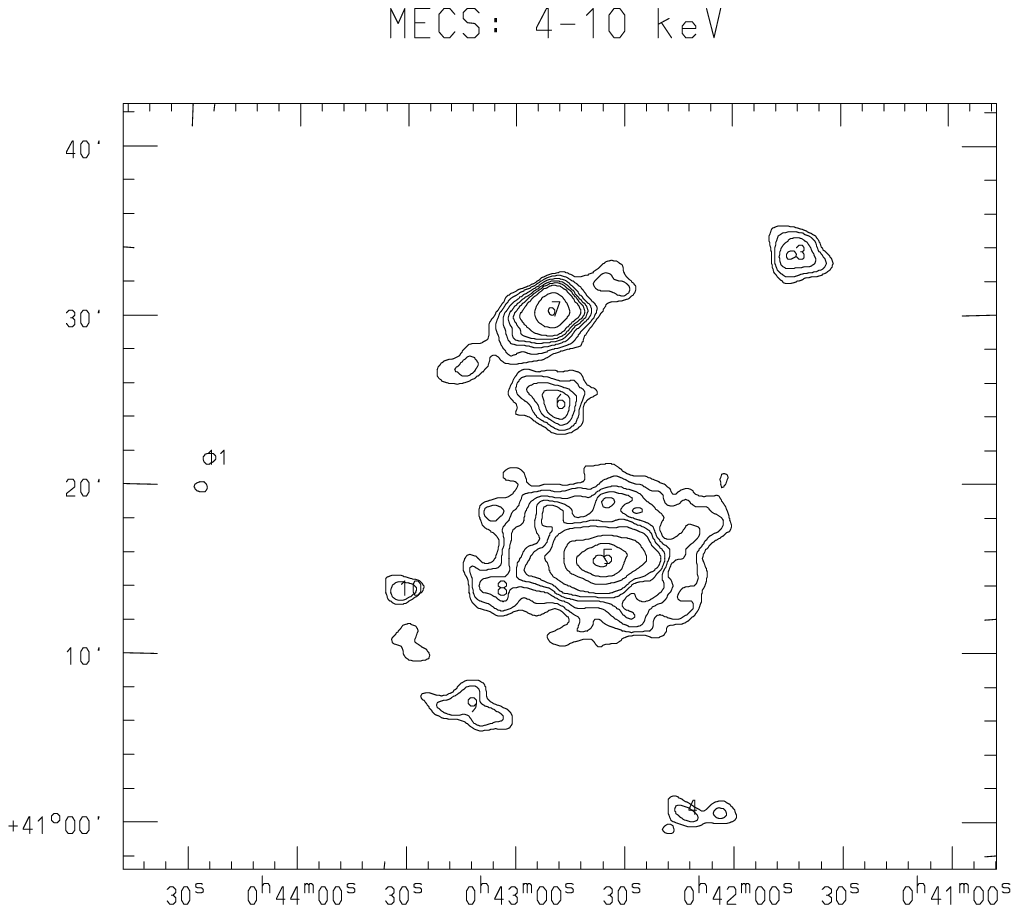,width=18cm,clip=}
}

\medskip
\resizebox{18cm}{!}{
\psfig{file=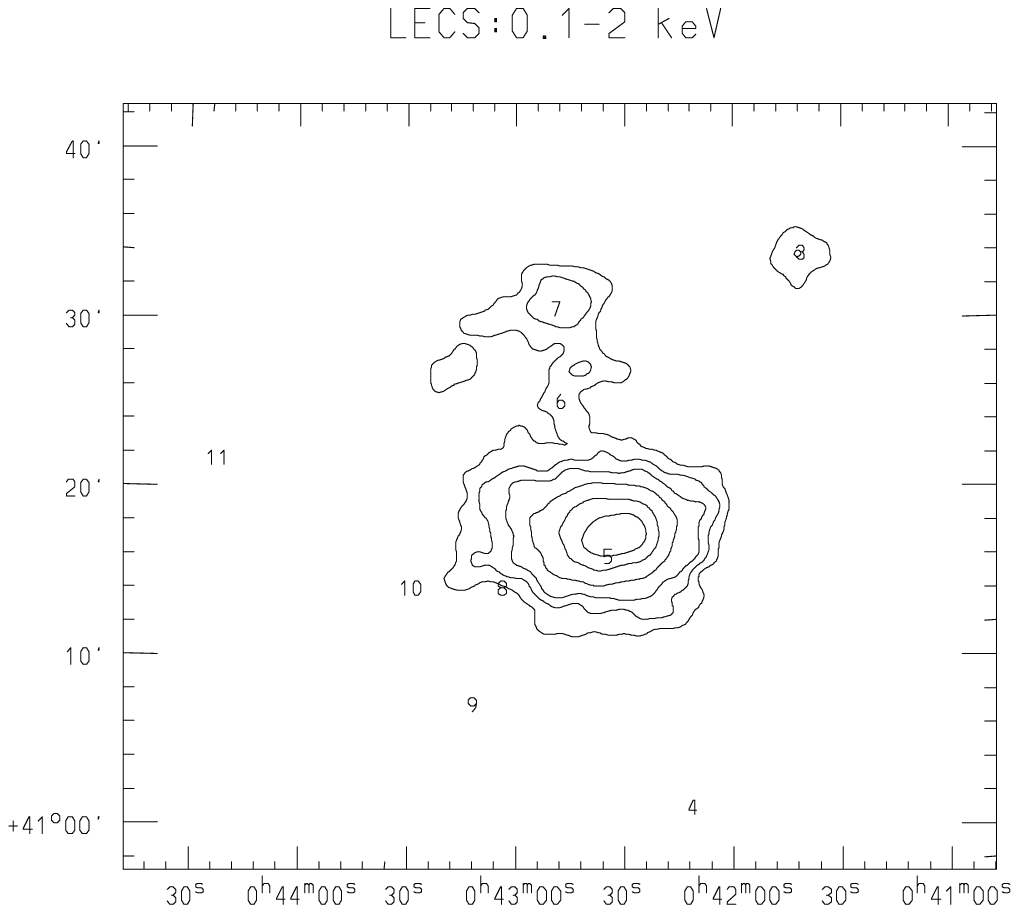,width=18cm,clip=}
\psfig{file=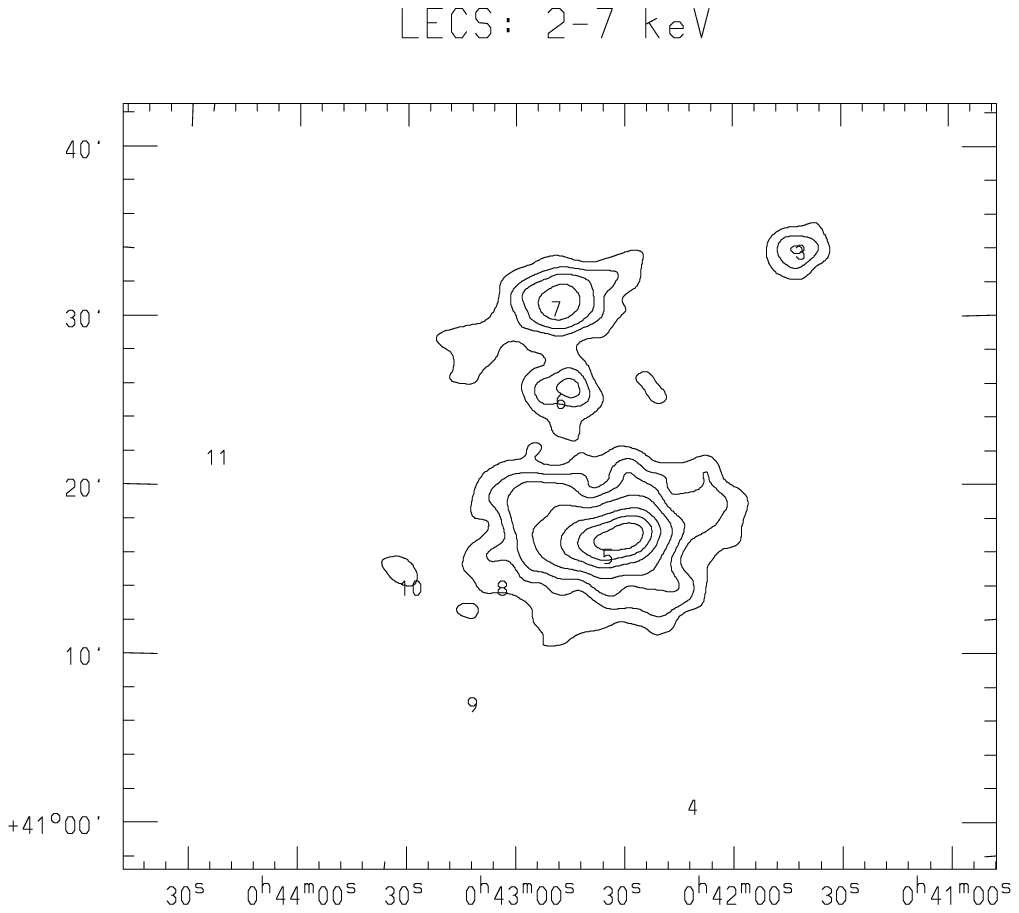,width=18cm,clip=}
}

\caption{Contour plot of the X--ray emission in M31 observed with \bsax
in field \#~3. TOP: MECS data, in two
different energy ranges: 1.8-10 keV (left) and 4-10 keV (right).
BOTTOM: LECS data, in two
different energy ranges: 0.1-2 keV (left) and 2-7 keV (right). 
The MECS data have been smoothed with a Gaussian function with $\sigma
=$ 24$''$, while $\sigma =$ 32$''$ is used for LECS data. 
The numbers indicate the positions of the sources' centroids
(identified with their numbers from Table~\ref{cross} in J2000 coordinates, 
determined from the 4-10 keV MECS data.  
Contours levels are: Upper left:  0.35 0.45
0.6 0.75 0.9 1.1 1.7 2.5 3.5 cnt/pixel; upper right:
0.55 0.7 0.9 1.1 1.5 2 2.5 3.5 5.5 cnt/pixel;  
Lower left: 0.1 0.18 0.28 0.5 0.9 1.3 1.7 cnt/pixel; 
Lower right: 0.1 0.18 0.28 0.5 0.9 1.3 1.7  cnt/pixel
}
\label{contoursf3}
\end{figure*}

At the distance of $\sim 700$ kpc, M31 is the normal, bright spiral
galaxy closest to us.  Moreover, it is also similar to the Milky Way in
size, metallicity and morphological type, and therefore can be used for
the dual purpose of investigating ${at~the~same~time}$ the properties
of our own and of more distant intermediate type spiral galaxies.  The
close proximity enables us to obtain very detailed observations in the
X--ray band also with current missions, and we can therefore study the
properties of its X--ray emitting evolved stellar population.  This
gives us the opportunity of better understanding analogous sources in
our own Galaxy.  There are several advantages of a detailed study of
M31 over our own Galaxy, in spite of the fact that sources are more
distant than Galactic objects, and therefore require higher
sensitivity and better spatial resolution: the distance to M31 is well
known, so that the luminosities of its sources can be accurately
calculated; the location of individual sources, $e.g.$ whether in the bulge or
in the disk of the galaxy, can be more easily assessed so that the
association with the stellar population is more reliable; the much
lower line--of--sight column density (N$_H$ $\sim$ 7 $\times
10^{20}$cm$^{-2}$ in our Galaxy) allows a more comprehensive
investigation of the spectral properties over a larger energy range
than it is possible in objects in the plane of our own Galaxy.
Moreover, due to its relatively favorable orientation, absorption
internal to the M31 disk is also reduced relative to that affecting
sources in the Milky Way disk.

\begin{table*}
\caption{Log of the MECS, LECS and PDS observations of the two fields on
M31}
\label{log}
\begin{tabular}{lccrccc}
\hline\hline
Name&R.A.&Dec.& begin--end&\multicolumn{3}{c}{Obs.Time (ks)$^1$}  \\
&\multicolumn{2}{c}{(J2000)}& &LECS&MECS&PDS \\
\hline\hline
Field \# 3 &0 42 29.45 &41 26 04 &22/12/97-24/12/97 & 38 & 88 & 39\\
Field \# 6 & 0 40 13.05 &40 50 10 &17/12/97-18/12/97 & 16 & 41  & 18 \\
\hline\hline
\end{tabular}

$^1$  Exposure times of LECS and PDS are shorter that those of MECS due
to the different observing modes of the three instruments.  MECS 
and PDS operate for all the useful observing time (with the exception
of  $\sim 5$ m. each orbit when the PDS instrument gain is calibrated
and the data are not used in scientific analysis).  However, because
of the collimator rocking, at any one moment only 2 out of 4 PDS units
are looking at the source, while the other two are used to estimate the
background, therefore giving $\le 1/2$ of the time on the source.  
LECS is operated only during satellite dark time, to
prevent contamination of the background by UV light entering the thin
organic window, significantly reducing the observing time.

\end{table*}

\begin{figure*}

\resizebox{18cm}{!}{
\psfig{file=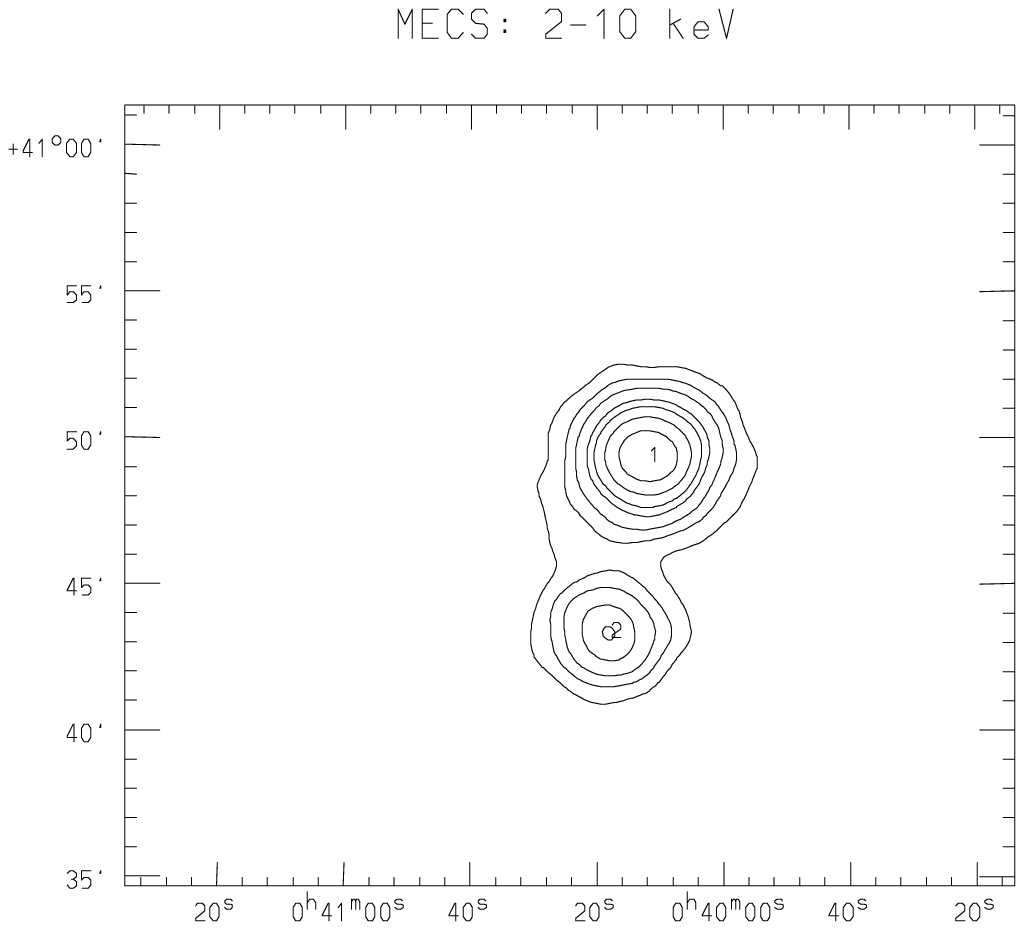,width=18cm,clip=}
\psfig{file=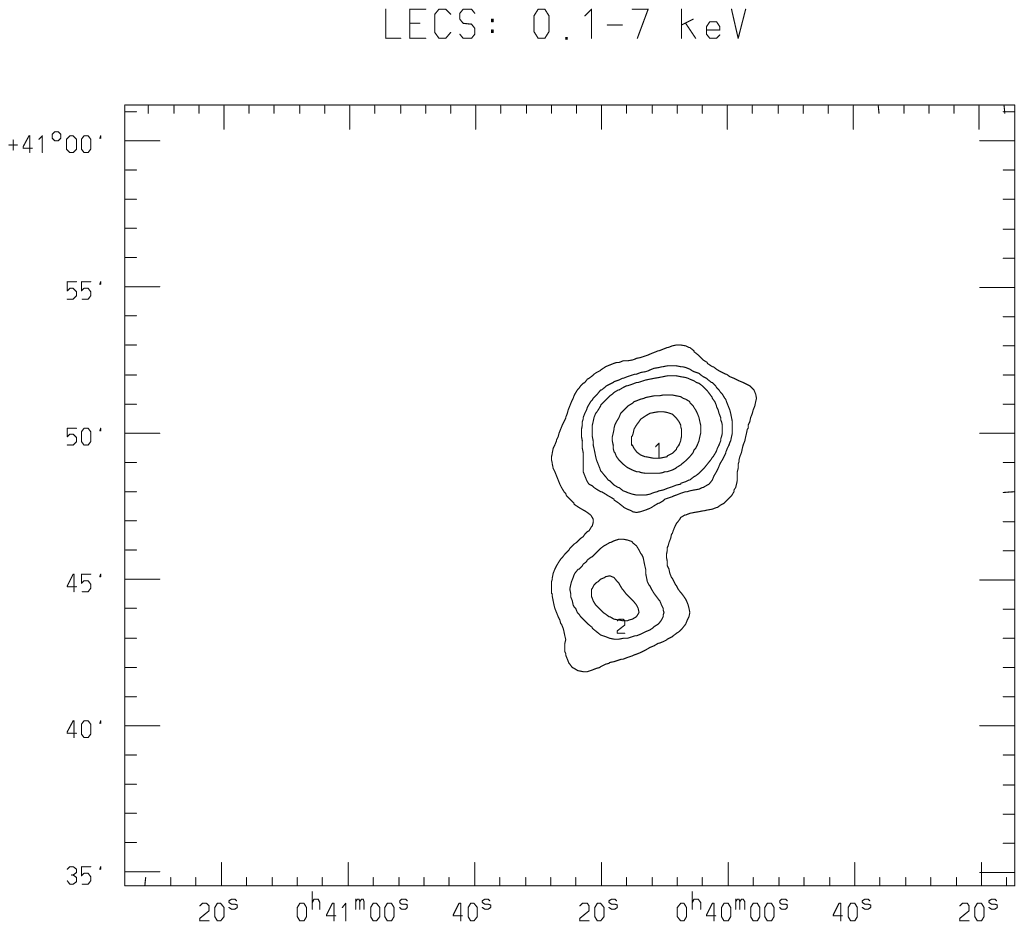,width=18cm,clip=}
}

\caption{Same as Fig.~\ref{contoursf3} for the observation of
field \#~6. LEFT: MECS data (1.8-10 keV);  RIGHT LECS data (0.1-7 keV). 
The data have been smoothed with a Gaussian function
with $\sigma =$ 24$''$ (MECS) and 32$''$ (LECS). 
Contour levels are: LEFT: 0.35 0.6 0.9 1.4 1.8 2.5 3.5 cnt/pixel;
RIGHT: 0.1 0.2 0.3 0.6 0.9 cnt/pixel
}
\label{contoursf6}
\end{figure*}

M31 has been the target of deep and detailed observations with all
previous and current X--ray missions.  Detailed maps have been obtained
in the soft energy band by $Einstein$ first and ROSAT more recently.
Over 100 sources were already detected with the $Einstein$
Observatory in the 0.2-4 keV energy band,
down to a luminosity of  10$^{36}$~erg s$^{-1}$ (Trinchieri \& Fabbiano
1991; TF hereafter).  ROSAT HRI and PSPC observations in the 0.1-2 keV
band have more than
tripled this number and have lowered the minimum detectable luminosity to a few
10$^{35}$~erg s$^{-1}$ (Supper et al.  1997, S97 hereafter; Primini et
al. 1993, P93  hereafter).

Several sources were detected in globular clusters and a few were found
associated with SNRs.   
A large fraction of the total
emission detected from M31 is concentrated in the bulge region,
where $\ge$ 50 sources have been individually detected (TF; 
P93). Some unresolved emission is also detected in the bulge.  P93
discuss that this is 
only in
part explainable with the integrated emission of faint unresolved
sources, while TF had attributed all of the emission to fainter
unresolved sources.  The overall integrated
X--ray emission was well fitted by a thermal bremsstrahlung
model with kT$\sim$6--13 keV testifying to the presence of very hard
X--ray sources (Fabbiano, Trinchieri and Van Speybroeck 1987). 

A detailed analysis of the spectral characteristics of single sources
has however remained largely unexplored so far.  IPC spectra were
obtained for a handful of them.  However, the limited statistical
significance of the detection, coupled with the limited spectral
capabilities of the instrument have given only tentative, and in some
cases puzzling results:  for example a higher value of the low energy cut--off
than expected on the basis of the total line-of-sight N$_H$ column density 
was observed in some of the sources identified with globular clusters.
The uncertainties on the characteristic temperatures were however so
large as to prevent any reliable conclusion on their spectral
characteristics.

ASCA has also obtained several pointings of M31 at higher energies, but
there are to date no reports in the literature of the results obtained.  
PSPC spectra of several globular clusters have been derived, however in
the limited and much softer ($\le$ 2 keV)
energy range provided by ROSAT (Irwin \& Bregman 1999).
We report here for the first time a study of the spectral properties of
the most luminous sources in M31 obtained with data from the \bsax
instruments, in the much wider $\sim 0.1-10$ keV band.  
We also analyze briefly the ASCA data for the bulge of M31,
to be compared with the \bsax results.

\section{Analysis of the BeppoSAX data}

\begin{figure*}

\resizebox{18cm}{!}{\hskip 8mm {
{MECS2 for Field \#~3 in detector coordinates ~~~~~~~~~~~~~~~~~~~~~~
MECS3 for Field \#~3 in detector coordinates }}
}
\medskip

\resizebox{18cm}{!}{
\psfig{file=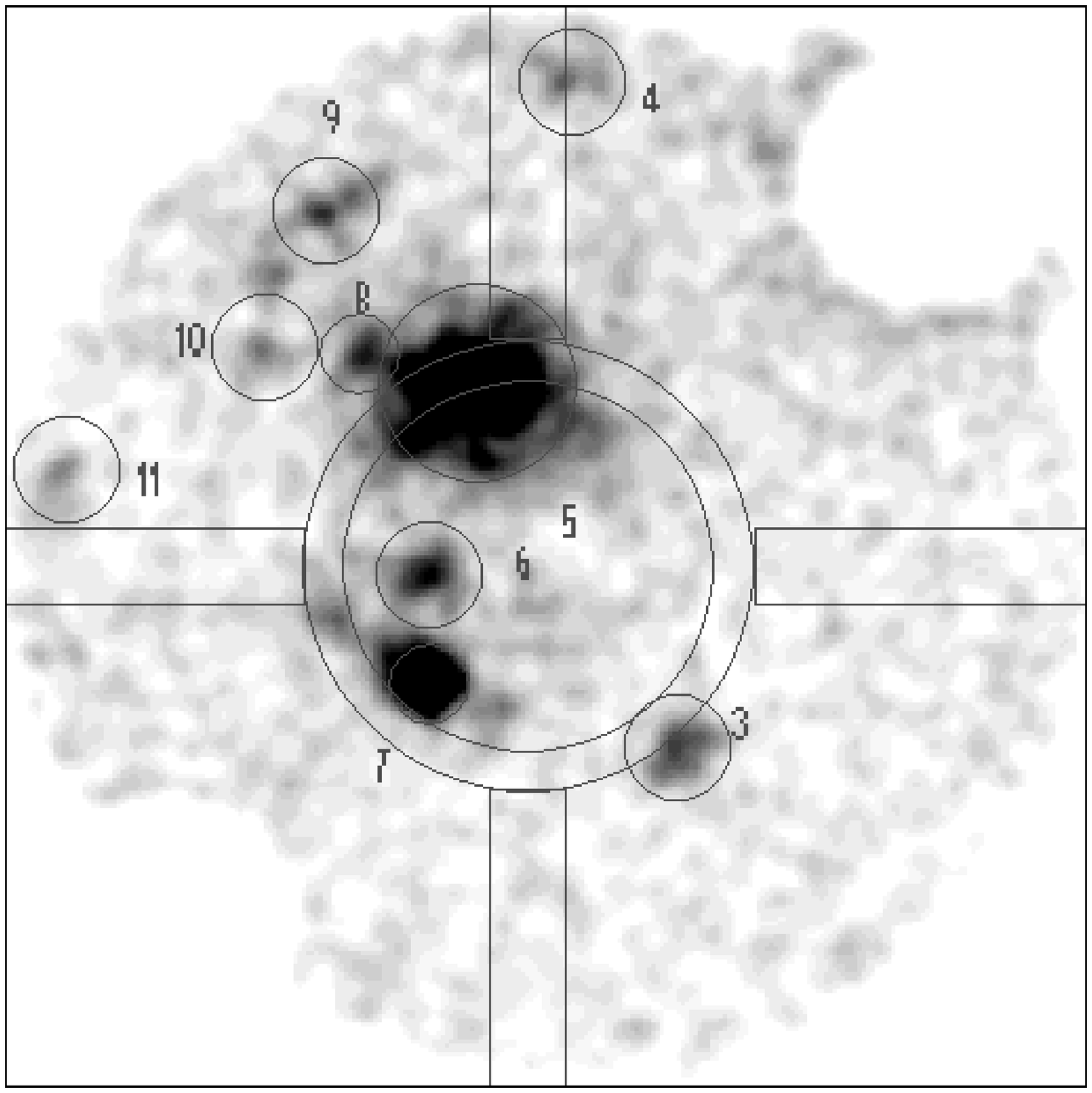,width=18.0cm,clip=}
\psfig{file=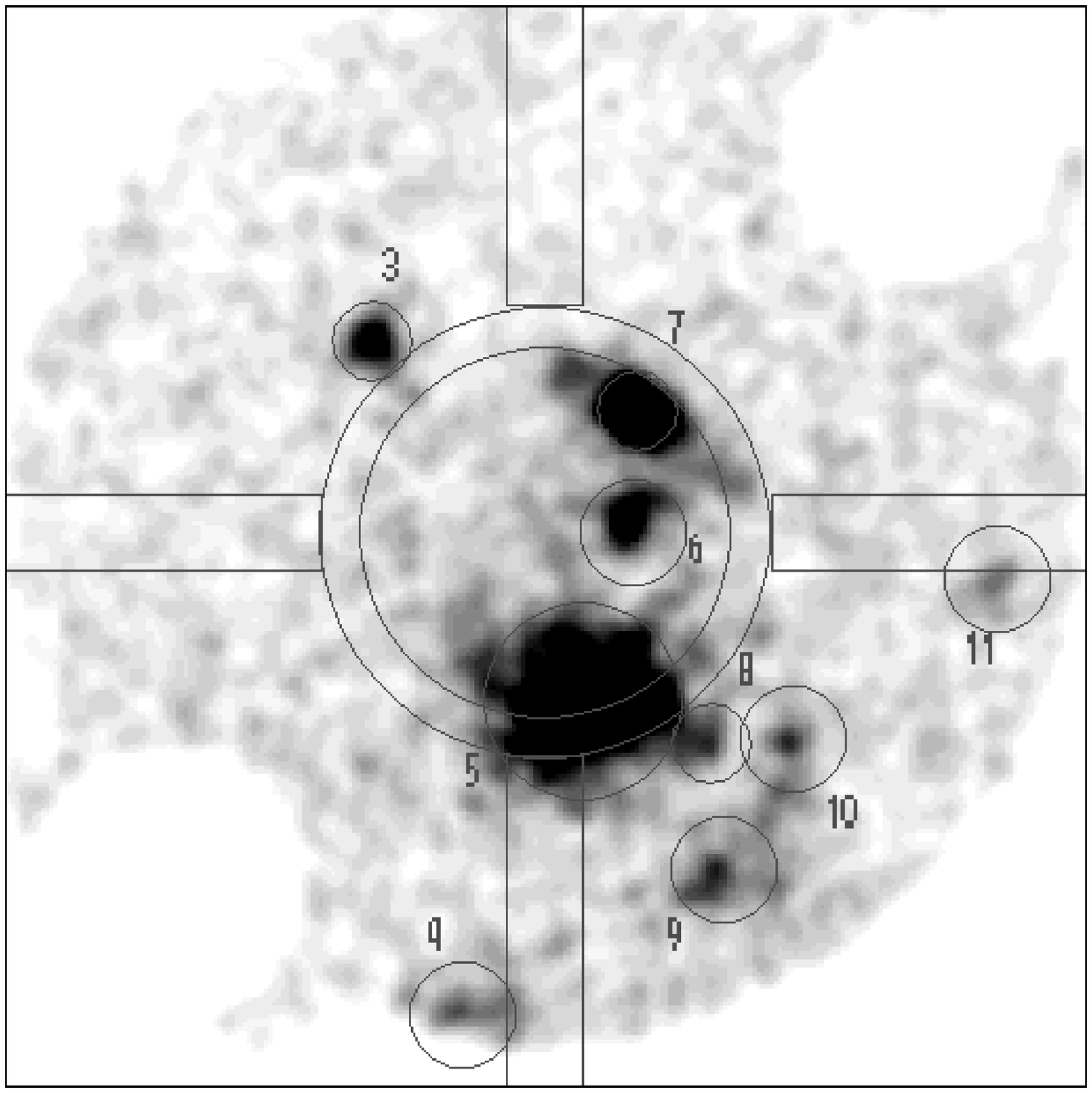,width=18.0cm,clip=}
}

\caption{The two MECS fields in detector coordinates for the
observation of Field \#~3.  Source positions
and detection cell sizes are shown, together with the rough position of 
the support structure.  This is schematized as a ring structure at
$\sim 9'-10'$ from the field's center, plus a cross-like structure
outside the ring (see Fig. 2 in Boella et al. 1997b). It should be
noted however that the extent of the support structure is not as
clear-cut as indicated in the figures.  The calibration sources are at
opposite corners (at the center of the ``white" circles in the upper 
left and lower right corners in the figure).
Note that MECS2 and MECS3 have opposite alignments relative to the
satellite axes. 
}
\label{strongback}
\end{figure*}

The X-ray astronomy satellite BeppoSAX (Satellite per Astronomia X,
named ``Beppo" in honor of Giuseppe Occhialini) is a Italian/Dutch
satellite developed, built and tested by a consortium of Italian and
Dutch Institutions, the Space Science Department of ESA and the Max
Planck Institut f\"ur extraterrestrische Physik.  The satellite and the
related instrumentation are presented in Butler \& Scarsi (1990),
Boella et al. (1997a) and references therein.  We present here the
observations of M31 obtained in December 1997 with 3 of the co-aligned
narrow field instruments: the Low Energy Concentrator Spectrometer
(LECS), sensitive between energies of 0.1 and 10 keV, with a circular
Field of View (FoV) of $ \sim 18.5'$ radius (Parmar et al. 1997); the
Medium Energy Concentrator Spectrometer (MECS), consisting of 2
identical active units sensitive between $\sim$ 1.3-10 keV and with a
FoV of $\sim 28'$ radius (Boella et al 1997b; a third unit was no longer
active at the time of our observations); and the Phoswich Detector
System (PDS), which is a non-imaging instruments composed of 4
independent units arranged in pairs (for on- and off-source
observations) sensitive in the $\sim 15-300$ keV band and with an
hexagonal FoV with FWHM $\sim 75'$ (Frontera et al. 1997 and
references therein).

In AO1 we obtained two pointings in the direction of M31: one (Field \#~3)
is centered north of the nucleus and contains the bulge, and a second
(Field \#~6) covers the SE region of the disk.  Table~\ref{log}
summarizes the parameters of the \bsax observations.

Figure~\ref{contoursf3}  and Fig.~\ref{contoursf6} show the X-ray
images obtained with \bsax (with the 2 MECS summed together and with
the LECS) in different energy bands.
As can be seen from the figures, several sources are detected in the
field.  Unfortunately, due to the configuration of the MECS instruments, and
considering that most of them are at large off-axis angles, several sources
are contaminated by
the support structure ($e.g.$ the ``strongback", see Boella et al. 1997b).  
Moreover,  they
fall onto different locations of the detectors,  as
illustrated schematically by Fig.~\ref{strongback} for Field \#~3, so
both the background and the contamination of the
support structure could be different in different instruments (the individual MECS units are
aligned differently with the satellite axes). 

\begin{figure}
\psfig{file=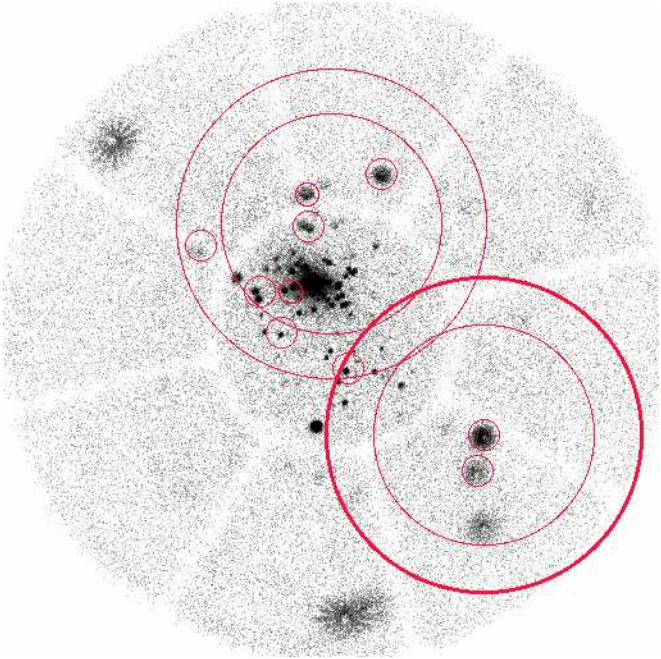,width=8.7cm,clip=}
\caption{Small circles of the size of the 
extraction regions for \bsax sources superposed onto a PSPC
observation of M31.  \bsax positions have been corrected by a constant
coordinate shift to better agree with the ROSAT coordinate system
(see text).
The MECS and LECS field of view are also illustrated with concentric
circles (larger one is for MECS) at the nominal pointing positions 
given in Table~\ref{log}.}
\label{pspcsax} 
\end{figure}

For a proper handling of the data, each individual
detector was analyzed separately, therefore the spectral distribution of
each source and the light curve
were derived separately, and then analyzed together, as explained
below. 
The pre-processed data provided by the Science Data Center (SDC), which
distributes cleaned and linearized event files in standard FITS OGIP
format, and the background files and response matrices (RMF) also
distributed by the SDC, have been used for the analysis.

\subsection{Sources in the \bsax fields and comparison with $Einstein$,
ROSAT and ASCA sources.}

Nine sources are detected in Field \#~3 and 3 in Field \#~6
(for a total of 11 source, since one is common to both fields)
with the MECS detectors of \bsax, as summarized in Table~\ref{cross}.  
Due to the much smaller observing
time, lower sensitivity and smaller field of view,
only sources~\#~1,2,3,5,6,7 have also been detected with the LECS.
No additional source is detected with LECS in either fields.  
Source~\#~5 coincides with the bulge of M31 and
is clearly extended/complex in the \bsax image, in agreement with the
clear detection of many sources with higher
resolution images from $Einstein$ [TF] or ROSAT [P93]).
All others are consistent with being single sources in \bsax, although 
more than one source could be present in the circle
used to determine fluxes and spectral parameters of these sources (see
Table~\ref{cross}).

\begin{table*}
\caption{M31 sources detected with the \bsax MECS and their proposed
identification with published lists in
the literature. }\label{cross}  
\begin{tabular}{lclrlrrrr}
\hline
\hline
Name&R.A.& DEC & R.A. & DEC& $Einstein$&\multicolumn{2}{c}{~~~ROSAT}& ident. \\
&&&\multicolumn{2}{c}{corrected}&&PSPC&HRI \\
&\multicolumn{4}{c}{(2000)}&\multicolumn{3}{c}{Source \#} \\
\hline
\hline
\multicolumn{7}{l}{Field \# 6:} \\
Source 1&00:40:11.6 & 40:49:22.0 & &  & 3 & 67 & &RS \\
Source 2&00:40:17.6 &  40:43:22.2&&  &4 & 73 && G \\
\multicolumn{7}{l}{Field \# 3:} \\
Source 3&00:41:42.4 &  41:33:39.4 &00:41:43.54 & 41:34:25.29 & 9 & 122 && G \\
Source 4&(00:42:12.5 &  41:00:51.9) &0:42:11.34&
        41:01:28.00& (16)&(138-139-150) &(13)&(G-No)\\
Source 5&00:42:35.8 &  41:15:40.0  &00:42:36.57&41:16:13.80& && &Bulge \\
Source 6&00:42:48.8 &  41:24:51.9&0:42:50.67&41:25:24.59  
       & 62 & 201-203 &53-59-61& No-SNR \\
Source 7&00:42:50.1& 41:30:19.9&00:42:52.51&41:30:53.07  & 67 & 205-207 &56 &
G-No  \\
Source 8&00:43:04.9 &  41:13:47.7&00:43:06.15&41:14:15.93  & 74-79-83 &
         214-220-225-228&65-68-70-74-76 & G-No \\
Source 9&00:43:13.1 &  41:06:53.9 &00:43:13.93&41:07:19.73 & 85& 229 &77& G\\
Source 10&00:43:31.9 &  41:13:47.0 &00:43:33.72&41:14:10.29 
       & 91-92 & 244-247&82-83& For-G\\
Source 11&00:44:25.4 &  41:21:28.4 &00:44:29.22&41:21:42.90 & 97 & 282 &&G \\
\hline
\hline
\end{tabular} 

\bigskip
NOTES: 

The corrected positions in Field \# 3 result from the plate solution using the
ROSAT positions (see text). \\
Source numbers are from Table 2 from TF, Table 5 from S97, and Table 1 from
P93 (note that the P93 list only covers sources \#4 to \#10). 
Optical identifications are also from Crampton et. al. (1984).  
G = globular cluster; RS = radio sources; SNR =
Supernova remnant;
For = foreground; No = no id \\
Source \#~4 is too close to the edge of the field for a reliable
determination of its position.  It is also in Field \#~6, 
at the very edge of the field and
close to the calibration source.  The position determined in the two
observations differ by $\sim 1^s$ in $\alpha$ and 2$'$ in $\delta$.
The $Einstein$ or ROSAT sources indicated could fall in the 2$\farcm6$
($2'$) circle used for the spectral analysis. \\
ROSAT sources \# 150 nd \# 205 are the more likely candidates for
sources \# 4 and \# 7 respectively (see text) 

\end{table*}

As already discussed, most of the sources are at large off-axis angles
in the detector, and in several cases fall near or under the support
structure of the instruments, which obscures photons at low energies
($\le 4$ keV).  This is particularly true for the bulge,
but also sources \#~3,\#~4, \#~7 and \#~11 could be affected by it, although at
different degrees of importance (see Fig.~\ref{strongback}).  This poses a
problem for determining both the spectrum and flux of these sources,
and their position.  In particular,
for source \#~5, the centroid determined in the 2-10 keV
band is $\alpha$=0:42:32.7 $\delta$=+41:16:04.0, while in the 4-10
keV band (where absorption from the strongback should be negligible)
this is $\alpha$=0:42:35.8 and $\delta$=+41:15:40.  
We have therefore determined centroids in the 4-10 keV band, where
contamination from the strongback should be negligible (LECS
positions do not appear to change significantly with energy). This is
also the band recommended by the SAX-SDC team, and it has been shown to
be reliable in comparison with the ROSAT data of the Marano field
(Giommi et al. 1998).

A further concern about source positions comes from the comparison of
LECS and MECS positions for the sources in common.  We find that
the absolute positions are not the same, but have a off-sets in the
range $\alpha \sim 16''-32''$ and 
$\delta \sim 40''-70'' $.   We estimate that 
$\sim 10''-15''$ is probably a reasonable assessment of the
average uncertainty in the determination of the peak position 
of sources in the MECS (larger for
very off-axis sources, also due to the asymmetry of the PSF at large
off-axis angles).  A similar uncertainty in the measure of
the aspect and misalignments between instruments and satellite axes, 
however, could approximately double the overall error.
We could therefore explain most of the discrepancies with a rigid
shift of the absolute 
coordinates. 

\begin{table*}
\caption{Results from the spectral fits to the MECS data. }\label{spectramecs}  
\begin{tabular}{lrrrrrll}
\hline
\hline
Name&MECS2&MECS3&$\Gamma$/kT& 90\% err& $\chi^2_\nu$ (DoF)&Unabs. Flux &Model/\\
&\multicolumn{2}{c}{cnt ks$^{-1}$ }&&&& (cgs) &Notes\\
\hline
\hline
Source 1& 26.59$\pm$0.82& 26.70$\pm$0.82& 1.94
&1.86-2.03& 1.1 (62)  & 4.7$\times10^{-12}$&P     \\
&&& 6.3 &5.6-7.3& 1.0 (62) &4.6$\times10^{-12}$&B\\
Source 2&10.69$\pm$0.53&10.02$\pm$0.53& 1.76
&1.62-1.90 & 1.3 (26)  & 2.2$\times10^{-12}$&P \\
&&&8.7 &6.5-12&1.1 (26)&2.2$\times10^{-12}$&B     \\
Source 3&  3.04$\pm$0.20 & 3.49$\pm$0.22
&1.41 & 1.23-1.60 & 0.8 (19) & 1.3$\times10^{-12}$&P \\
&&& 28.9 & 13-160& 0.8 (19) &1.3$\times10^{-12}$&B \\
Source 4& 2.41$\pm$0.20& ---&
0.80&0.43-1.12&1.0 (7)&2.5$\times10^{-12}$&P $^b$\\
&&&200&$>43$&1.7 (7)&2.0$\times10^{-12}$&B $^{b,c}$\\
Source 5& 61.00$\pm$0.86& 50.69$\pm$0.79& 
&&&& $^d$\\
&28.11$\pm$0.59&25.46$\pm$0.57&
2.24&2.12-2.35&1.6 (41)&1.8$\times10^{-11}$&P\\ 
&&&6.1&5.4-6.2&1.3 (41)&1.7$\times10^{-11}$&B \\
Source 6&5.72$\pm$0.29 &  5.60$\pm$0.28 &
1.74 & 1.59-1.88 & 0.7 (34)  & 1.2$\times10^{-12}$&P\\
&&& 9.7 & 7-15& 0.8 (34)& 1.2$\times10^{-12}$&B \\
Source 7& 11.29$\pm$0.37& 13.75$\pm$0.41& 
1.70 &1.61-1.77 & 1.2 (64) &3.7$\times10^{-12}$&P$^a$ \\
&&&10.0& 8-12& 1.0 (64)  & 3.6$\times10^{-12}$&B\\
Source 8& 4.09$\pm$0.24 & 3.70$\pm$0.23& 1.82
& 1.66-2.00 & 1.3 (22)  & 1.4$\times10^{-12}$&P\\
&&& 7.7 & 5.6-12& 1.3 (22) & 1.4$\times10^{-12}$&B \\
Source 9& 3.38$\pm$0.22 & 3.56$\pm$0.23&      
1.05&0.83-1.21&1.2 (22)&2.4$\times10^{-12}$&P \\
&&& 200 & $>$60 &1.3 (22)&2.2$\times10^{-12}$&B$^c$     \\
Source 10& 2.79$\pm$0.22 & 3.33$\pm$0.23
& 1.86&1.66-2.07 & 1.2 (20)  & 1.4$\times10^{-12}$&P$^a$\\
&&&6.3 &4.5-9.7&1.0 (20)&1.3$\times10^{-12}$&B \\
Source 11&1.76$\pm$0.17  &2.15$\pm$0.20
&1.87&1.55-2.2&0.8 (13) &1.3$\times10^{-12}$& P \\
&&&7.6&4.3-20  &1.0 (13)&1.2$\times10^{-12}$&B\\

\hline
\hline
\end{tabular} 

\bigskip
NOTES:  
sources 1 and 2 are in field \#~6, sources 3 to   11 in field \#~3. 
P stands for Power Law model, and B for Bremsstrahlung. 
kT  is in keV. $\Gamma$ is the photon index.   Net observing
times are 87906 s. for MECS2 and 87777 s. for MECS3 in field \#~3, and
41609 (MECS2) and 41424 (MECS3) in field \#~6.
Fluxes are the average value between the two MECS in the 2-10 keV range.

\begin{description}
\item $^a$ A broken power law provides a better fit to these data.  Best
fit parameters are:  
$\Gamma_1$ = 1.5 , E$_B$ = 5.4, $\Gamma_2$ = 2.8 ($\chi^2_\nu$ = 1 for 62
DoF) for source 7, with a F-test probability of $> 99.99$; 
$\Gamma_1$ = 1.5 , E$_B$
= 4, $\Gamma_2$ = 2.9  ($\chi^2_\nu$ = 1 for 18 DoF) for source 10
(F-test probability $\sim 99.1$). 
\item $^b$ Source 4 is at the border in MECS3.
\item $^c$ Best fit value is hard pegged at the maximum allowed 
value in the fit.  
\item $^d$ No reasonable fit can be obtained using the full energy
range.  Energies above 3.5 keV only are considered in the next two
lines.
\end{description}

\end{table*}

For the purpose of the cross identification of sources, we have used
MECS positions, that are available for all sources, and we have
compared them with published source lists (TF; S97; P93).  We have
first identified the \bsax sources with the closest $Einstein$ and
ROSAT source(s) to the positions in Table~\ref{cross}.   We find that
all \bsax sources have both an $Einstein$ and a ROSAT counterpart.
For 5 sources, confusion is not an issue: there
is only one $Einstein$ and one ROSAT source as the possible
identification of the \bsax sources. Moreover, the position of source
\#7 is very close to that of the ROSAT PSPC 
source \# 205.  We have then compared the
\bsax and ROSAT PSPC positions and found a systematic negative shift in
declination, of an amplitude in the range 35$''-44''$ for most sources,
and a less clear pattern in R.A. (mostly a negative shift of 1-2 sec)
with the closest identification.  Using the 4 sources in Field \# 3
with unique ROSAT PSPC counterparts as reference celestial coordinates,
we could indeed find a different set of coordinates for the \bsax
sources, with a RMS astrometric error of $\sim 10''$.  The newly
determined coordinates differ by an average 1.5$^s$ in R.A. and $\sim
30''$ in declination from the old ones, although not by a constant shift
equal for all sources.  This is consistent with a possible $\le 1'$
systematic offset in the absolute \bsax\ positions.  

We have checked again the cross-identifications between the 
\bsax\ sources and published lists, either using the newly
determined coordinates (however valid only for Field \#3) or
equivalently applying the average shift to the coordinates in
Table~\ref{cross}, that can be done for all sources.   Table~\ref{cross}
lists as possible identification all sources that would be included in
the circle used for the spectral analysis (see next section), in spite
of the fact that they might not be the most likely identification,
either because they are farther from the expected position (for
example, the position of \bsax sources \#~7 would be at $\ge
1'$ from ROSAT source \#~207) and/or because
much fainter than other candidates.  Sources \#~1, \#~2, \#~3, \#~9 and
\#~11 are identified with one source only.   PSPC
ROSAT sources \# 150 and \# 205 are the most likely candidates of \bsax\
sources \#4 and \# 7 respectively.  
Sources  \#6 and \# 10 have more than 1 ROSAT counterpart  (2
$Einstein$ sources for \#~10) that could contribute equally to the
\bsax\ fluxes.  The spectral results should therefore be treated with
caution, since they could be the superposition of intrinsically
different spectra.   Source \#8 is very close to the confused
bulge area.

\begin{figure*}
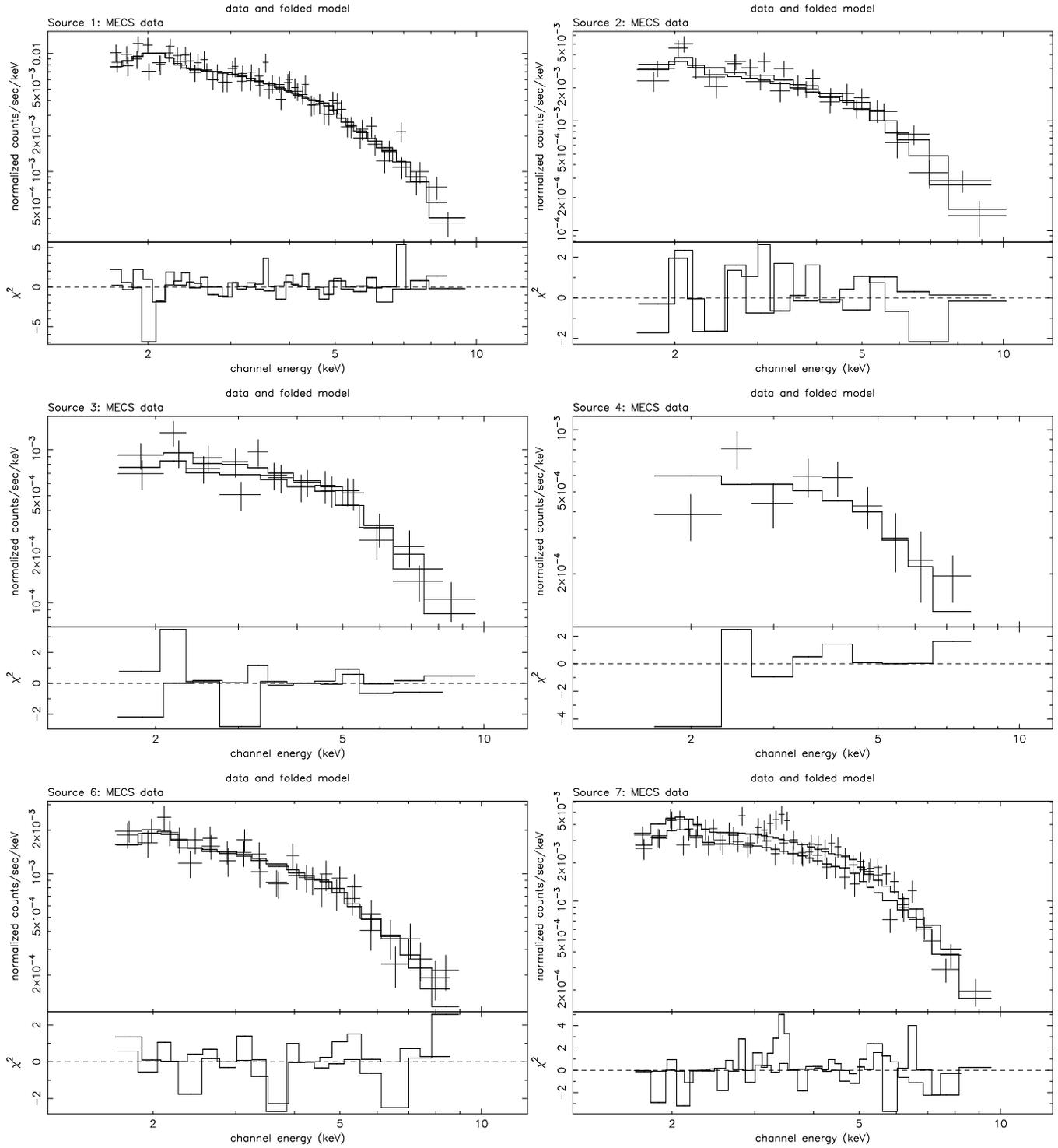

\resizebox{18cm}{!}{
\psfig{file=8757f5a.ps,width=18cm,angle=-90,clip=}
\psfig{file=8757f5b.ps,width=18cm,angle=-90,clip=}
}

\medskip
\resizebox{18cm}{!}{
\psfig{file=8757f5c.ps,width=18cm,angle=-90,clip=}
\psfig{file=8757f5d.ps,width=18cm,angle=-90,clip=}
}

\medskip
\resizebox{18cm}{!}{
\psfig{file=8757f5e.ps,width=18cm,angle=-90,clip=}
\psfig{file=8757f5f.ps,width=18cm,angle=-90,clip=}
}

\bigskip
\caption{Spectral distribution of the sources detected with 
the 2 MECS instruments and $\chi^2$ distribution, assuming a  
B model to fit the data.}
\label{mespec}
\end{figure*}

\begin{figure*}
\addtocounter{figure}{-1}
\medskip
\resizebox{18cm}{!}{
\psfig{file=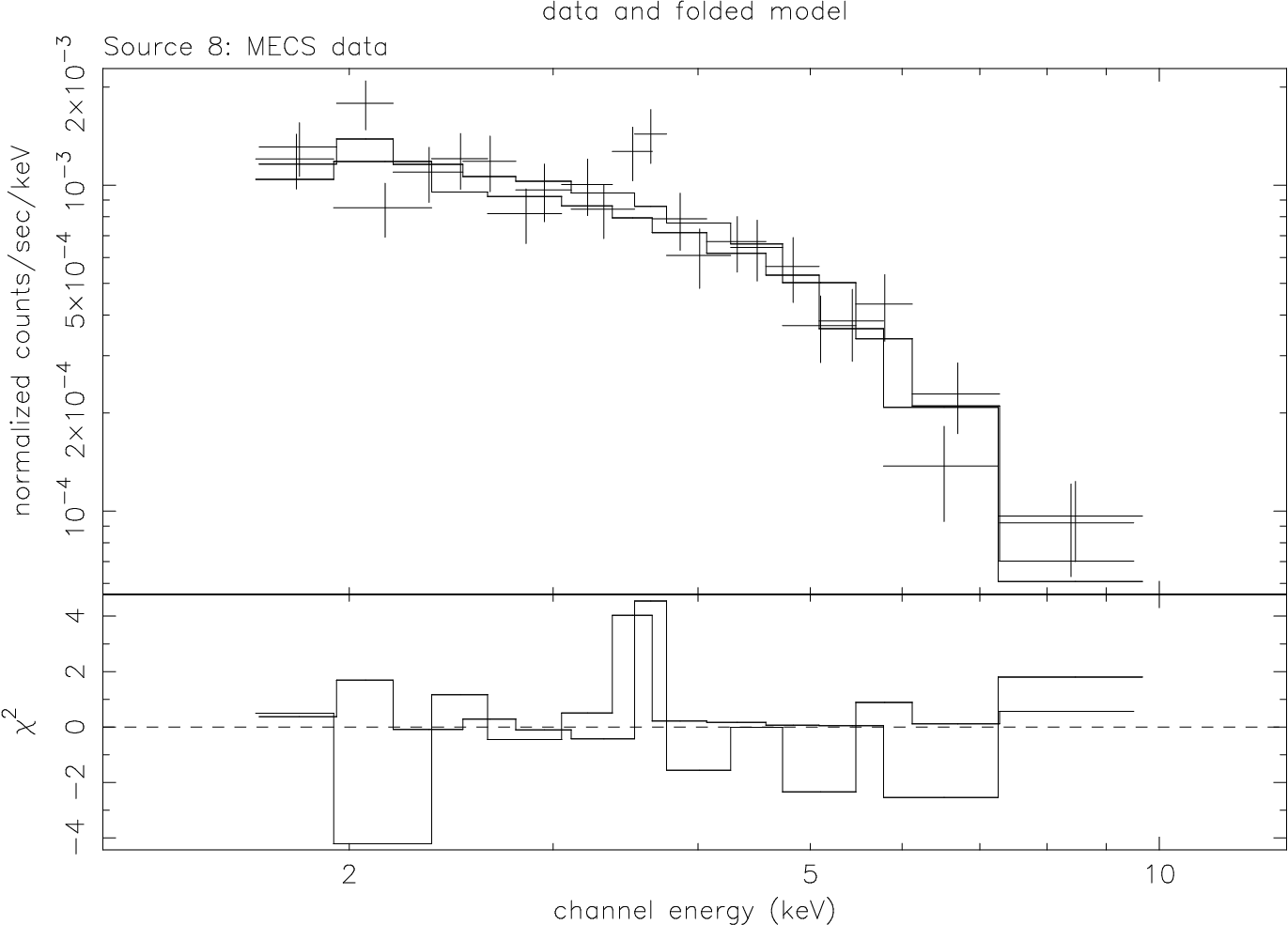,width=18cm,angle=-90,clip=}
\psfig{file=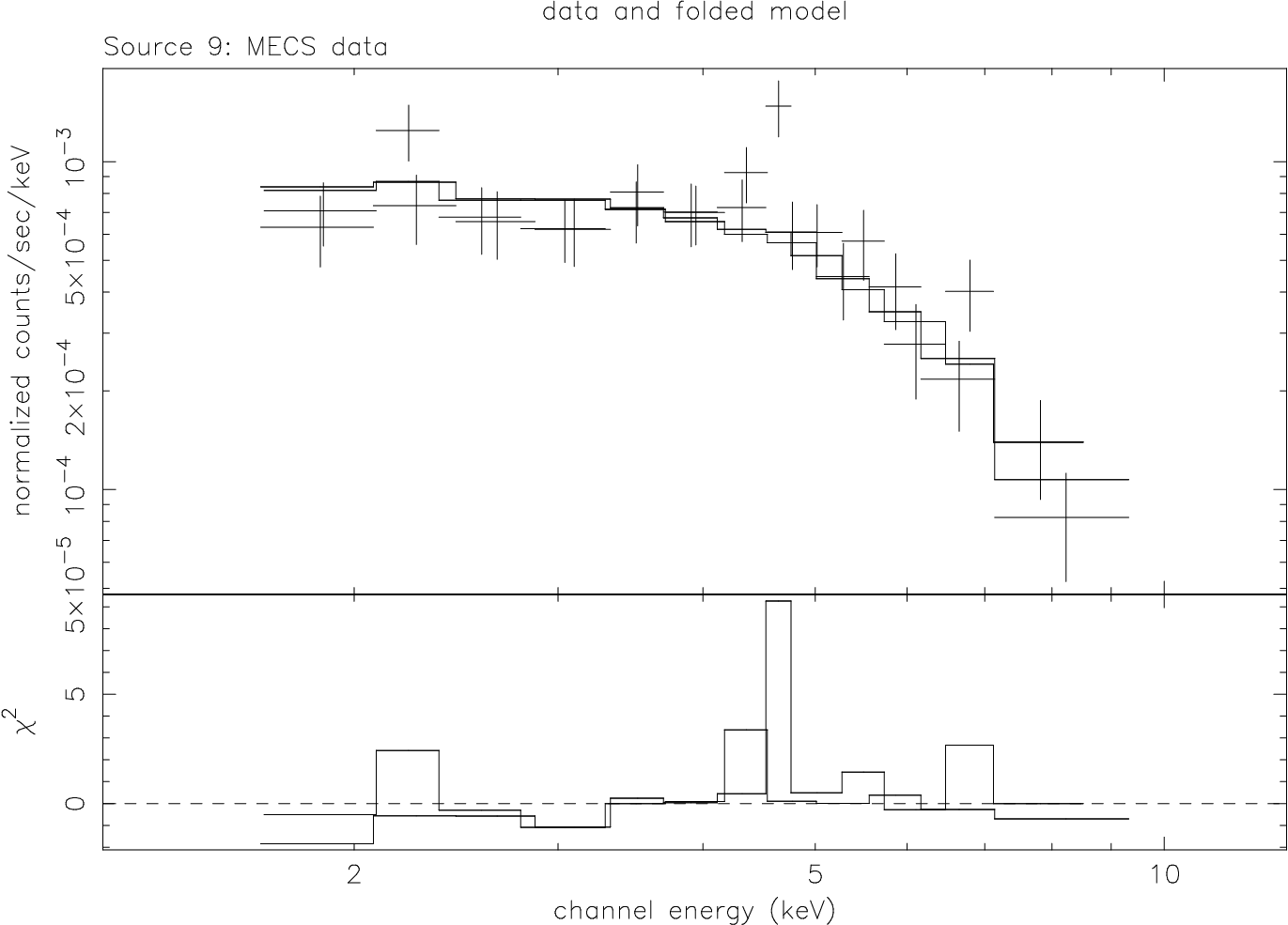,width=18cm,angle=-90,clip=}
}

\medskip
\resizebox{18cm}{!}{
\psfig{file=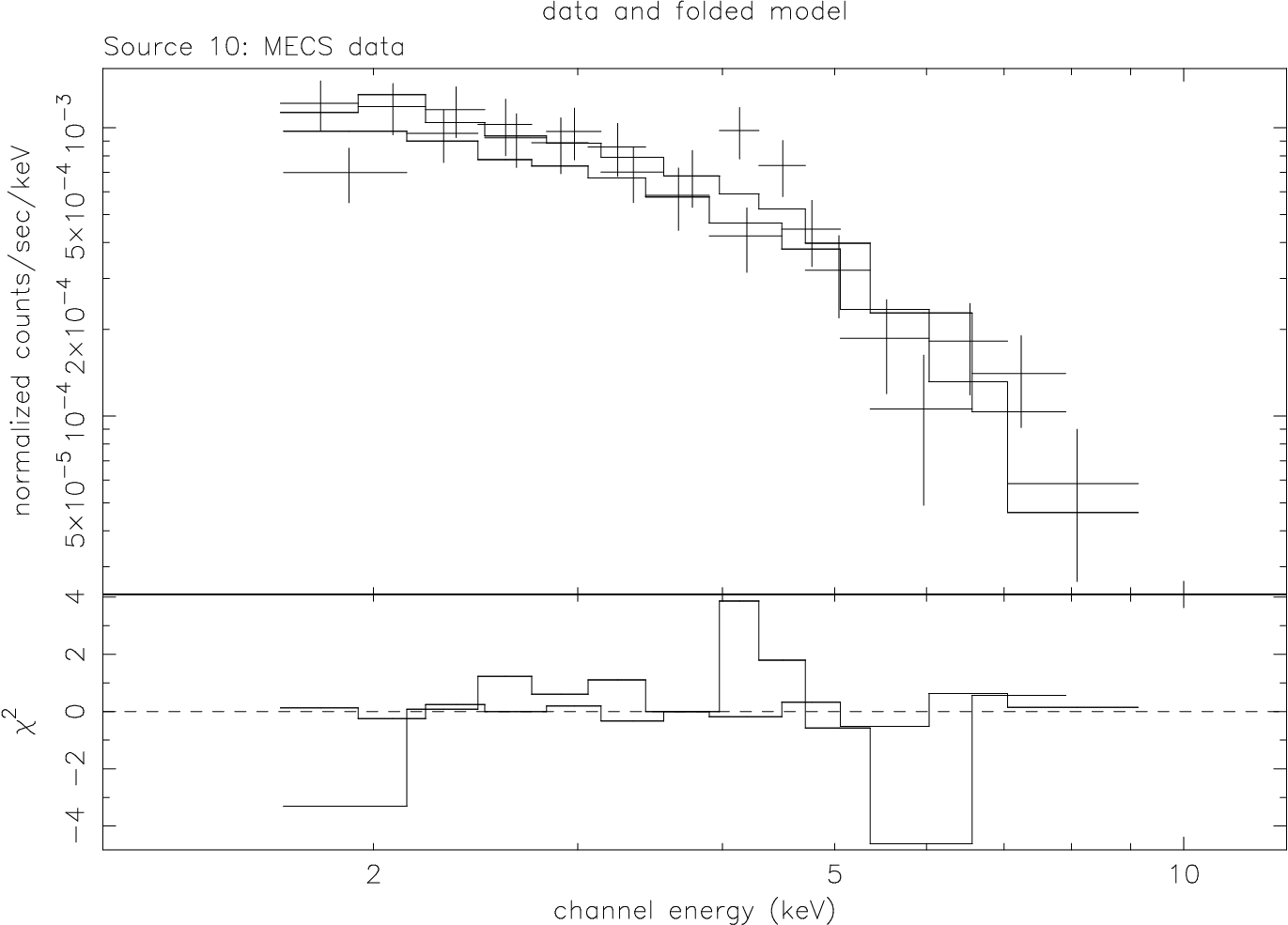,width=18cm,angle=-90,clip=}
\psfig{file=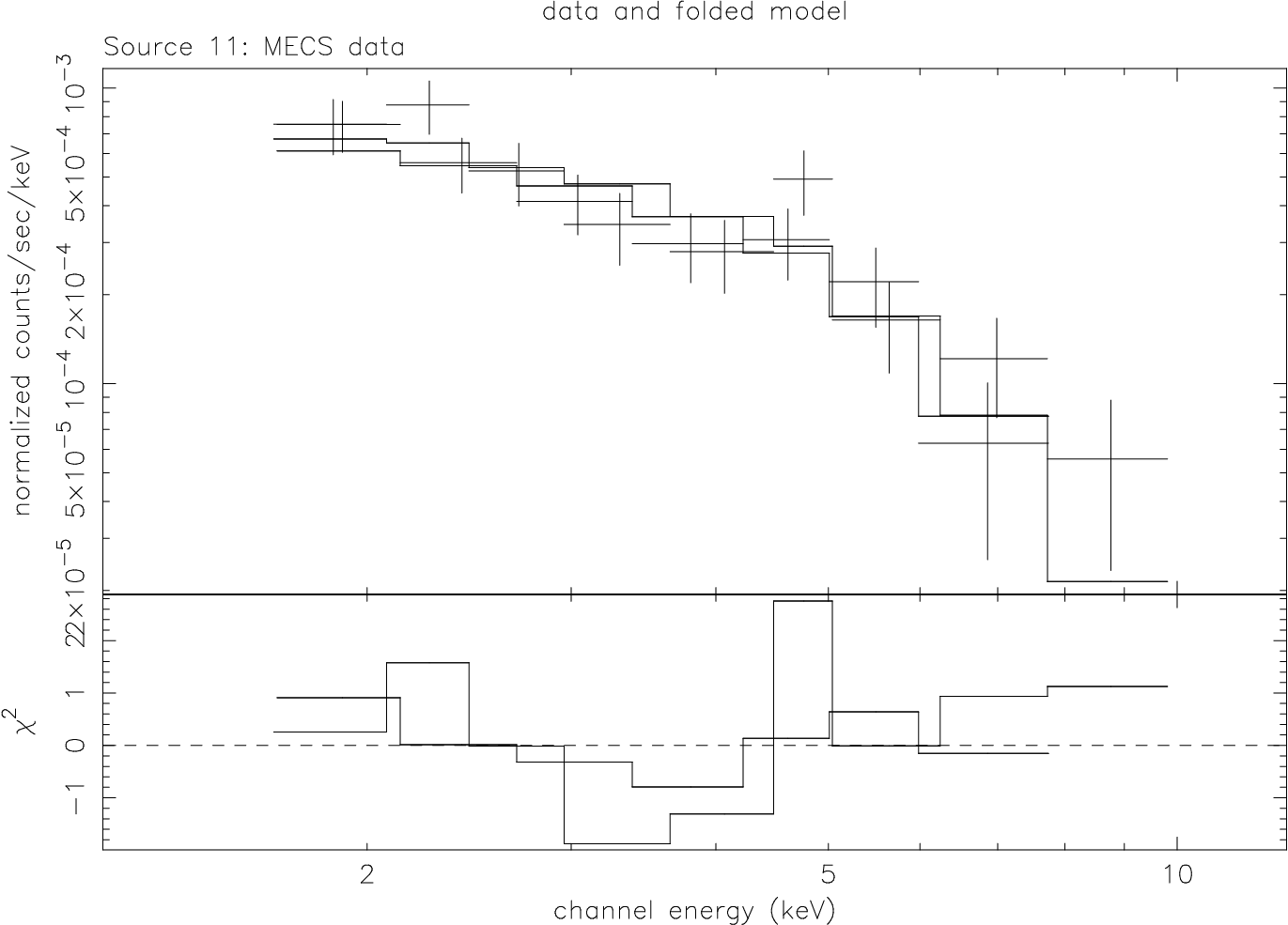,width=18cm,angle=-90,clip=}
}

\bigskip
\caption{{\sl--Continued}}

\end{figure*}

When identified, the proposed counterparts of the X-ray sources are
for the vast majority  globular clusters (Crampton et al. 1984, S97, P93). 
Source \#~1 (and
possibly also source\#~10) could be unrelated to the galaxy (see
identification list in Crampton et al. 1984).

\subsection{Spectra of individual sources: MECS data}

Since most sources are expected to be point like, the determination of
the spectral distribution of the source photons should be relatively
straightforward.  However, as discussed above,  contamination from the
support structure is heavy (see Fig.~\ref{strongback});   
moreover, the field is crowded, so we
cannot use the standard  $\sim 4'$ detection cells for these sources either
because of overlap or because of their vicinity to the strongback.  We
have therefore resolved to use a fixed detection cell of 2$\farcm$6
radius for all sources except \#~5 (the bulge) and \#~7 and \#~8, for
which the radii were 5$'$, 2$'$ and 2$'$ respectively, and to center
the cells in each instrument  using the 4-10 keV image 
so as to minimize the influence
caused by differential absorption due to the support structure. 
The ``Area Response file" (ARF) for the chosen cell size at the
appropriate off-axis and azimuthal angles was obtained for each source
in each detector using the program $accumulate~matrix$ in the XAS
software environment distributed by the SDC,   which also includes a
correction for the presence of the ``strongback".  The resulting ARF
computes the fraction of PSF included within the source extraction
radius using the numeric model of the on-axis PSF, which is calibrated
within 6$'$, but that has been verified to be valid out to off-axis
angles of 10$'$ (Molendi, private communication).  Moreover, the
spectral analysis of two well known sources (4U0142+61 and RX
J0146.9+6121) at different off-axis angles further testifies to the
reliability of the matrixes produced even outside of the ``official"
calibration region (Israel et al. 1999; Mereghetti et al., in prep.).

Table~\ref{spectramecs} summarizes the count rates, and best fit parameters
for each source obtained with XSPEC.  
To produce these numbers we have first selected the
regions around the sources (plotted on
Fig.~\ref{strongback}).  Given the different positions of the sources in
the 2 detectors, we have analyzed
each MECS separately, in order to properly assess the expected
background at the detector position and possibly deal with different
covering fractions from the ``strongback".  Although small, the chosen cell
size should be large enough that small variations in the centering of
the cells in the two detectors do not introduce significant
differences in the count rate of each source.  The relative
normalization, which is a free parameter in the fit to ensure that
small residual differences in the efficiencies of the two detectors are
taken into account, should also correct for errors that might arise from the
possibly different covering of the total flux from each source
(see also Fiore et al. 1999).   The spectral
distribution of each source has been extracted with XSELECT,
and the data have then been rebinned to improve on the signal-to-noise,
typically to have a minimum of 30 total counts in each channel.  
Channels at energies below $\sim 1.8$ keV and above 10 keV are not
considered.  The background is estimated from the same detector position
from the corresponding  background event files.

For the spectral fits, we 
have assumed either a  power law or a bremsstrahlung spectrum with  
the line-of-sight column density fixed at 7$\times 10^{20}$
cm$^{-2}$. More
sophisticated spectra are not required, since in most cases one or
either of the two models we used approximates well the spectral
distribution of the photons.  Moreover, the limited statistics of the
detection does not allow us to properly test models with more
parameters. In a few cases where the minimum reduced $\chi^2$
($\chi^2_\nu$)
value prefers one model over the other, we have also tried different 
fits (namely a broken power law, see Table~\ref{spectramecs}).  
We have imposed the same $\Gamma$ or kT
for both instruments, but let the normalization between the
two instruments as a free parameter.  The relative normalization is
typically $\le$ 10\%, but it becomes $>$ 10\% for sources \#~7 and \#~8,
which could be an indication of different degrees of contamination from
the support structure or the neighboring source, and/or of a different
centering precision  in the two instruments.   For source \#~5 the
relative normalizations differ by as much as factors $>$2, become
closer to $\sim$40\% for energies above 3.5 keV and to $\sim$ 12\%
above 5 keV.

\begin{figure*}
\resizebox{18cm}{!}{
\psfig{file=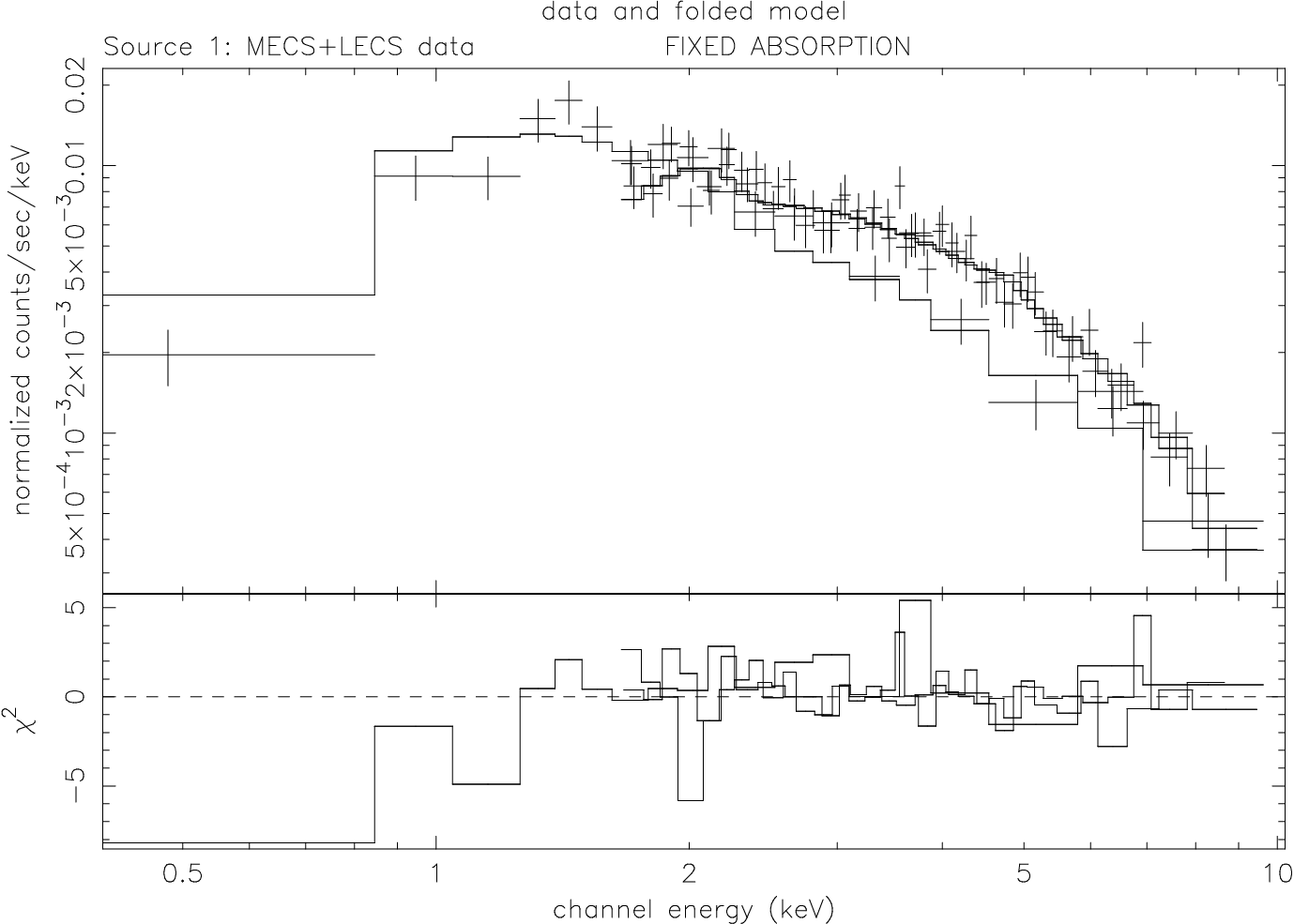,width=18cm,angle=-90,clip=}
\psfig{file=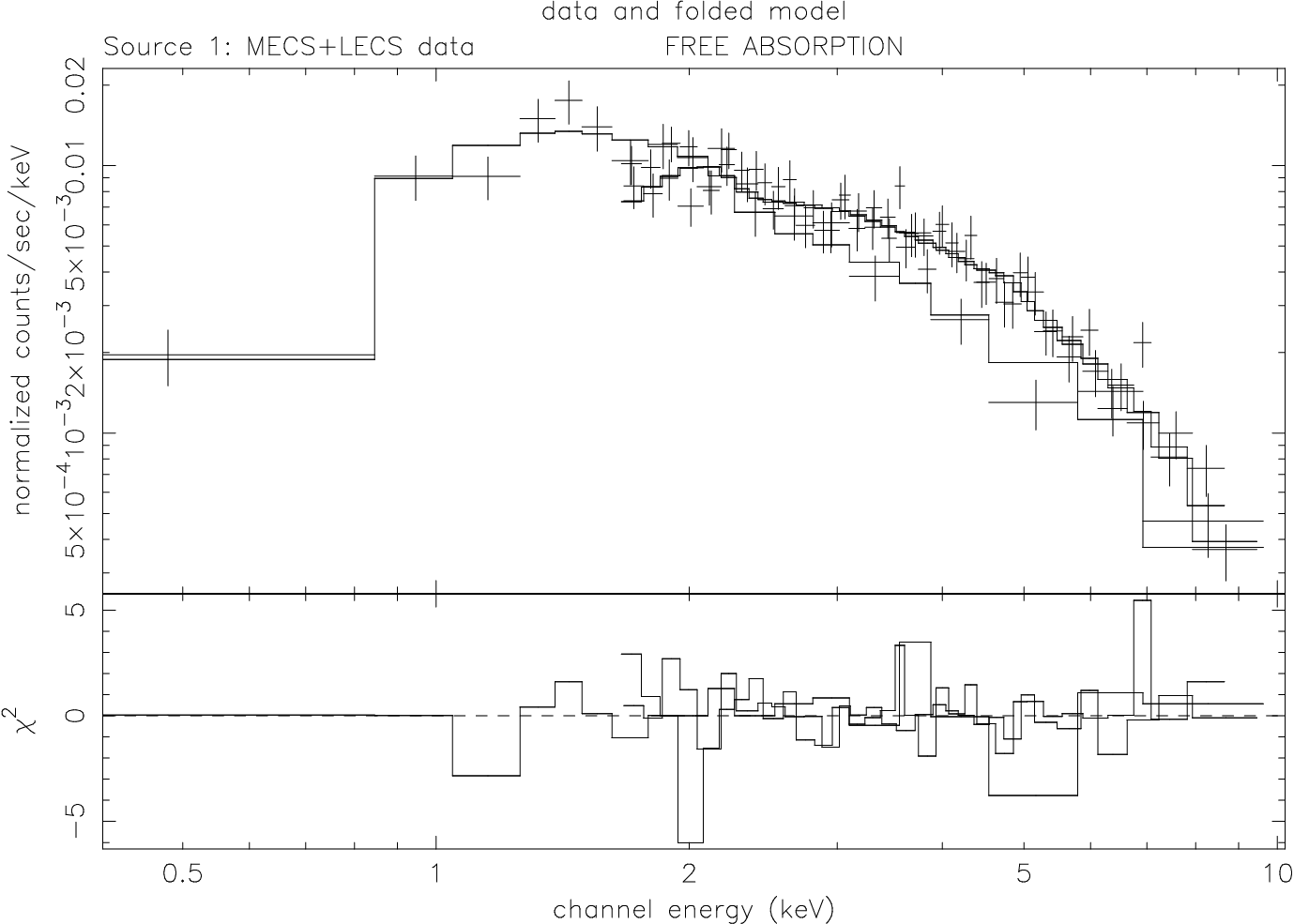,width=18cm,angle=-90,clip=}
}

\medskip
\resizebox{18cm}{!}{
\psfig{file=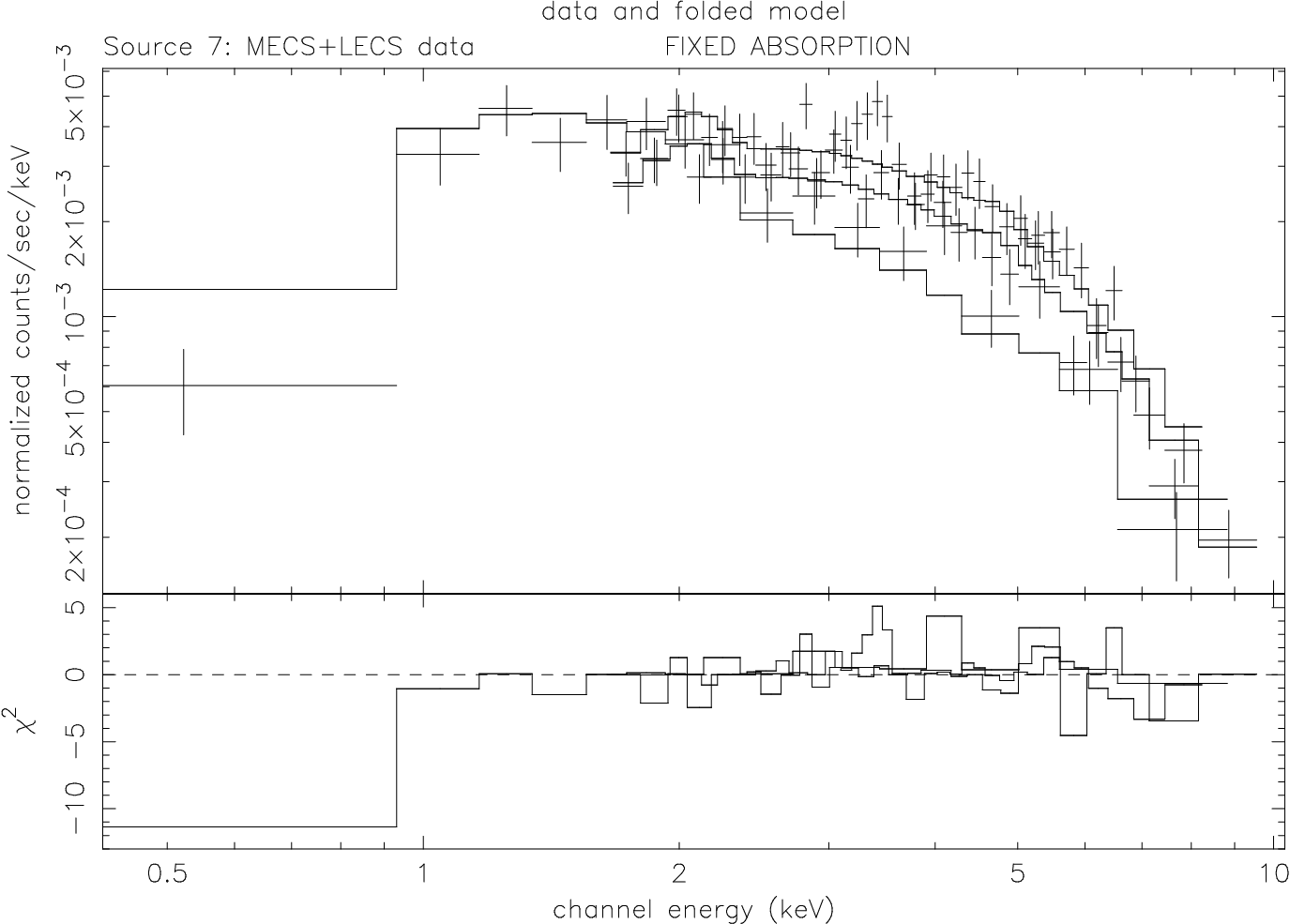,width=18cm,angle=-90,clip=}
\psfig{file=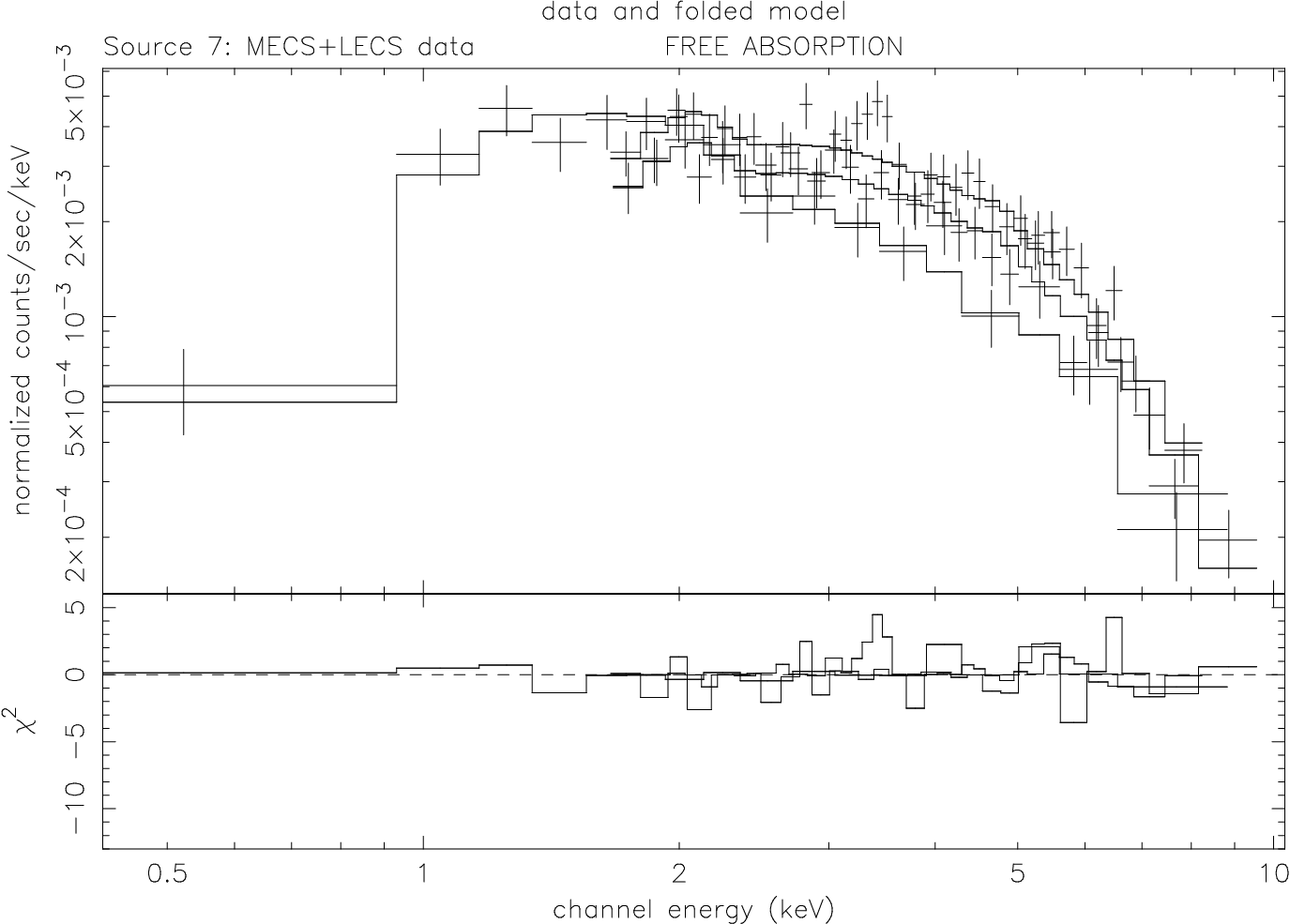,width=18cm,angle=-90,clip=}
}

\caption{Spectral distribution of the photons in the sources common to
MECS and LECS.  Source \#~1 and \#~7, for which a higher than
line-of-sight absorption is suggested, are plotted twice, 
with the absorption model parameter  N$_H$=7 \e20
cm$^{-2}$ (LEFT) and   N$_H$ at the best fit value in
Table~\ref{lecs+mecs}
(RIGHT).}
\label{melespec}
\end{figure*}

As shown by Table~\ref{spectramecs}, in most cases either model is
adequate, and the best fit parameters are $\Gamma \sim 1.8$ or kT $\sim
6-10$ keV.  The $\chi^2$ is smaller for the B model for sources \#~2,
\#~7
and \#~10.  This could indicate a preference over the P model, as also
suggested by the fact that a broken power law indeed lowers the minimum
$\chi^2$ value (see Table~\ref{spectramecs}), indicating that a model
with curvature is preferable.  The relatively high values of the
minimum $\chi^2_\nu$ for sources \#~8 and \#~9 are mostly due to a couple
of bins that strongly deviate from the model prediction.  However, they
are most likely statistical fluctuations, since the residuals do not
show systematic deviations from the mean (they are seen in one
instrument only, or there are large positive residuals balanced by
similar negative ones), so they should not be used as a strong
indication of a poor fit (see Fig.~\ref{mespec}).

\begin{figure}
\addtocounter{figure}{-1}

\psfig{file=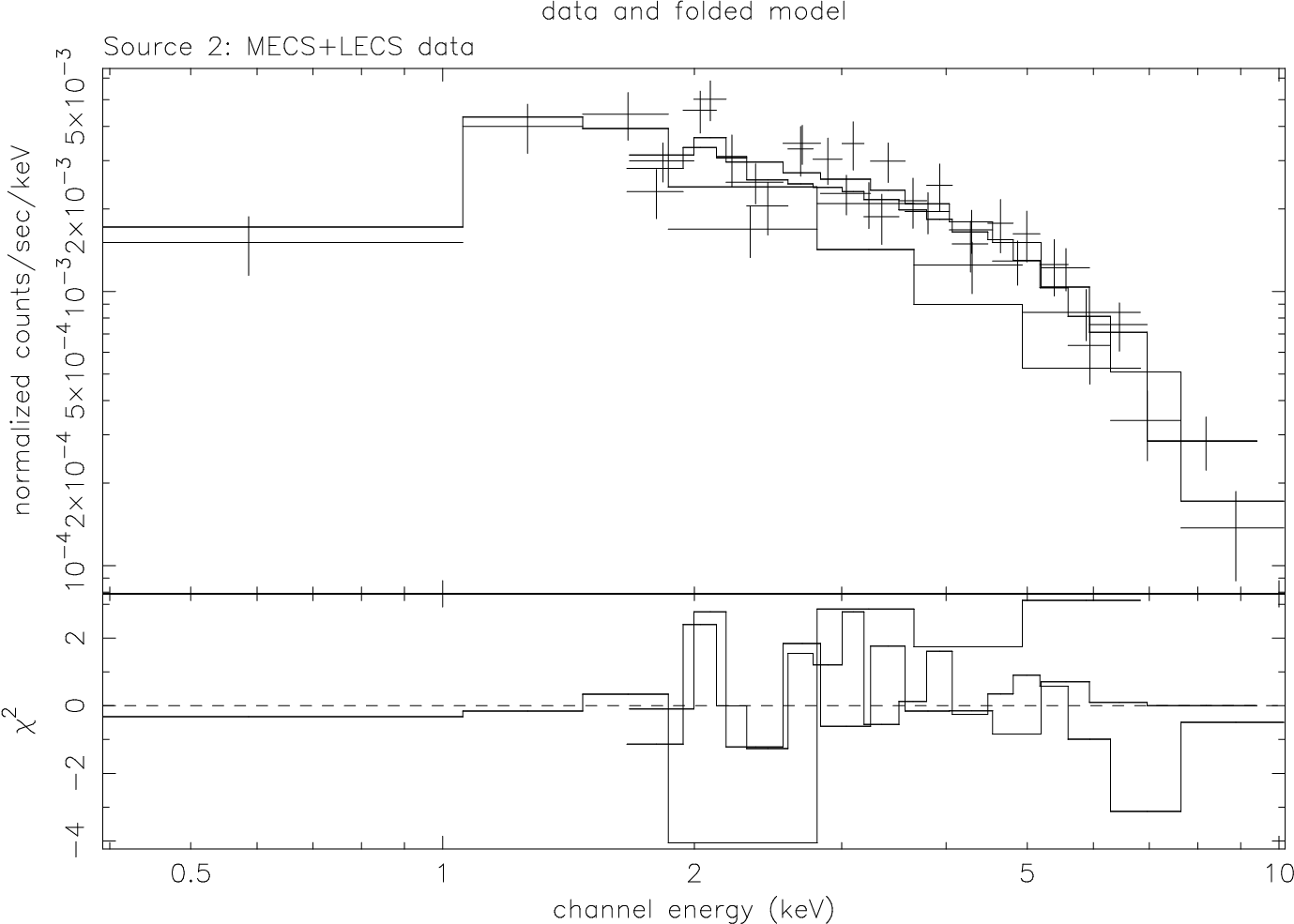,width=8.5cm,angle=-90,clip=}
\medskip

\psfig{file=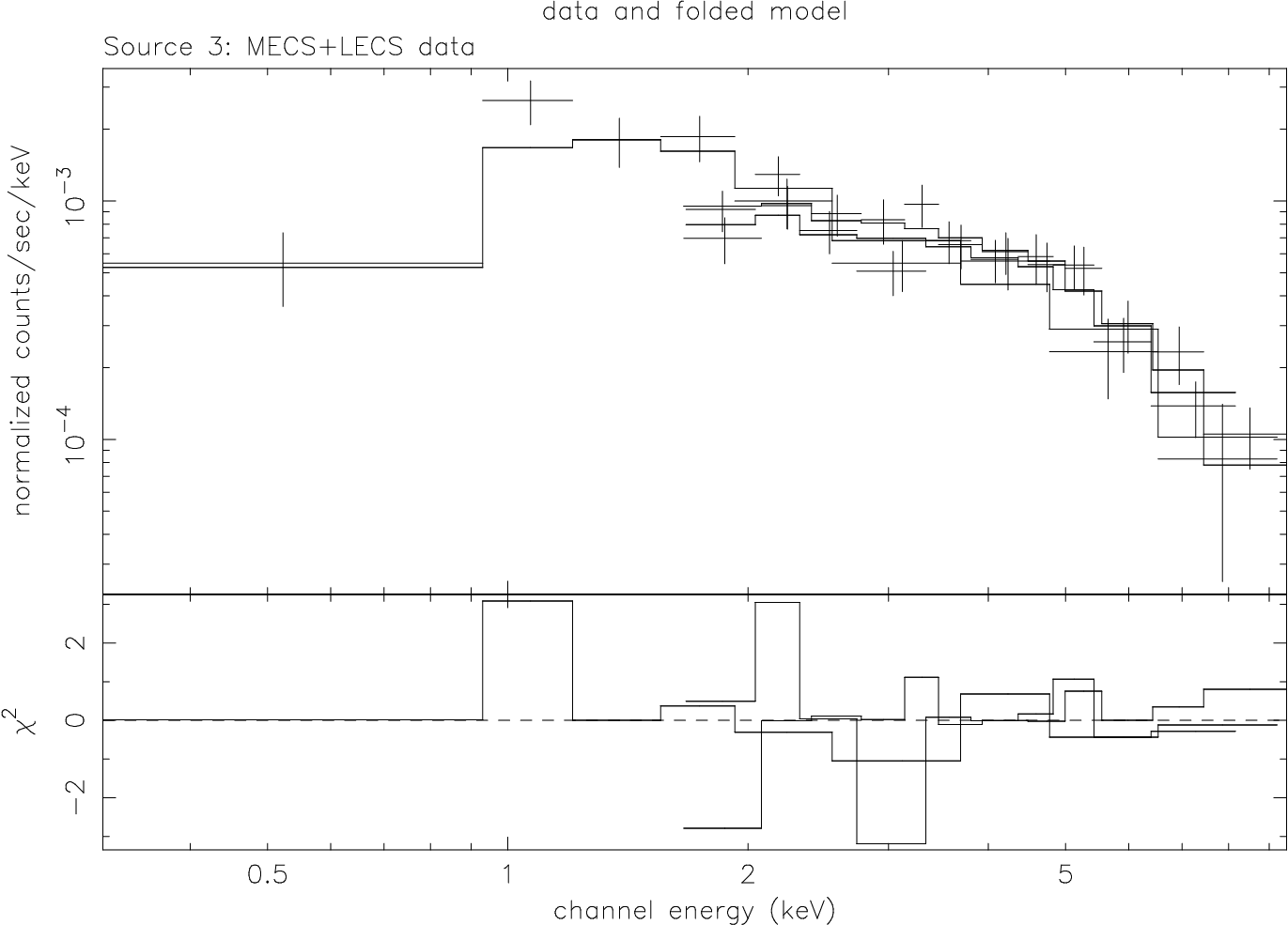,width=8.5cm,angle=-90,clip=}
\medskip

\psfig{file=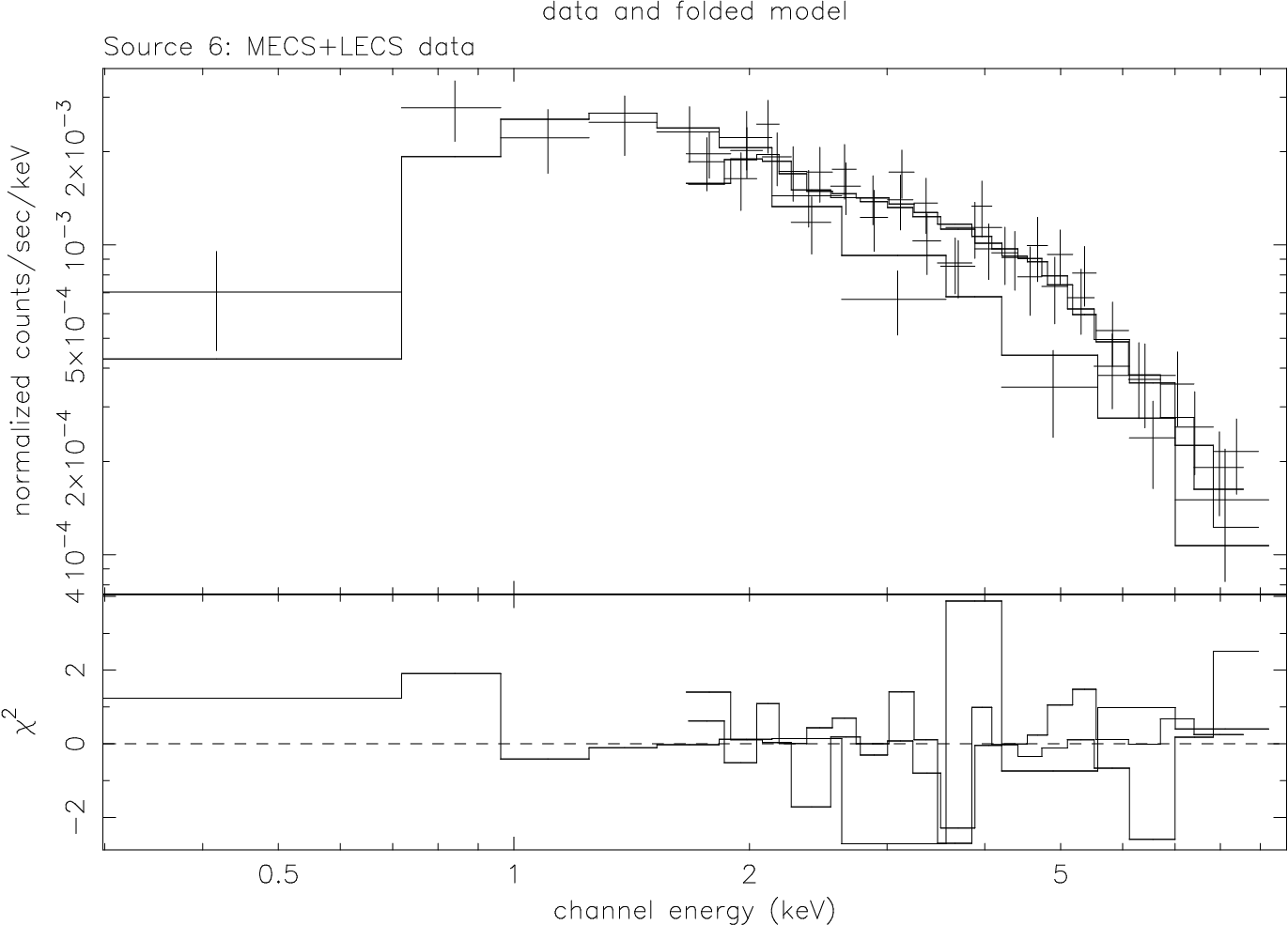,width=8.5cm,angle=-90,clip=}

\caption{{\sl --Continued }}

\end{figure}

Three sources (\#~3, \#~4, \#~9) have
significantly different best fit parameters from the others.    The
results for source \#~4, which is at the edge of the field, should
probably not be regarded as significant, since calibration at such
extreme off-axis angles is not reliable. 
Sources \#~3 and \#~9 are significantly harder than the others.  We have
tried to understand whether their different spectrum could be due to
spurious effects. 
Source \#~3 could be influenced by the support structure,
although the effective area file should have taken this into account.
We have nonetheless excluded photons below 4 keV from the fit and found
very similar best fit values, although with clearly much larger errors.  
Source \#~9 should not be affected by the strongback, and in this case
too a fit to high energy photons only reproduces the best fit parameters
listed in Table~\ref{spectramecs}.  We have also tried the standard
4$'$ detection cell,
which is possible since there are no neighboring sources nor the 
support structure, to investigate whether we have assumed too small a
detection cell for the instrument PSF (although this should have
affected other sources as well, and should be taken into accounts by the
ARF), and once again found consistent best fit parameters. 
The release of the constrain on N$_H$ does not alter significantly the
best fit values either. 
We therefore believe that these two sources are significantly harder than
the other sources in M31.

The bulge of M31 (source \#~5)
cannot be fitted with either of the models considered
above, when the full range of energies are considered.  In
Table~\ref{spectramecs} we include the bulge results for completeness, 
but we give the spectral parameters derived from the
data at high energies only, for which a fit could be obtained with the 
models used for all other sources.  Given the heavy obscuration from the
strongback, discarding all low energy photons is a safe procedure. 
A more detailed treatment of the spectral data for the bulge is given
in \S~\ref{thebulge}.

\subsection{Spectra of individual sources: LECS+MECS data}

The significantly smaller observing time obtained with the LECS causes a
much poorer detection efficiency in this instrument. Moreover, we have
used only a fraction of the detected counts for each source, since the
field is crowded and we cannot adopt the standard detection cell of 8$'$
radius that would ensure that 95\% of 0.28 keV photons are included
within the selected area.  We have used detection cells of the same
size as those used for the MECS, centered at the peak position as seen
by the LECS.  We have produced effective area files with
the $lemat$ program in the SAXDAS software environment distributed by
the SDC.  The mirror response and strongback obscuration are modeled by
means of ray-tracing (see Parmar et al. 1997).

\begin{table*}
\caption{Results from the joint spectral fits to the MECS and LECS 
data. }\label{lecs+mecs}  
\begin{tabular}{lrrrrrrllll}
\hline
\hline
Name&LECS & $\Gamma$/kT& 90\% err& N$_H$& 90\% err&$\chi^2_\nu$ (DoF)
&\multicolumn{3}{c}{Unabsorbed
Flux (cgs)} &Model/\\
&counts& &&&&&2-10 keV&0.1-2 keV& 0.2-4 keV &Notes\\
\hline
\hline
Source 1&542$\pm 24$&  2.08 &1.92-2.20&50&38-67&1.0 (79)  
&3.4$\times10^{-12}$&7.1$\times10^{-12}$& 6.9$\times10^{-12}$&P     \\
&& 5.8 &5.0-7.0&26&18-40&1.0 (79) 
&3.4$\times10^{-12}$&3.0$\times10^{-12}$&4.4$\times10^{-12}$&B\\
Source 2&186$\pm15$& 1.78 &1.60-2.00 &30&13-60& 1.5 (31)  &
1.7$\times10^{-12}$&1.9$\times10^{-12}$& 2.2$\times10^{-12}$&P $^a$  \\
&&10.2&7.6-15& 7&--&1.3 (32)
&1.6$\times10^{-12}$&1.0$\times10^{-12}$&1.5$\times10^{-12}$&B $^a$     \\
Source 3&198$\pm17$& 1.50 & 1.33-1.63 & 7&--& 0.8 (27) 
&1.1$\times10^{-12}$&6.8$\times10^{-13}$& 9.7$\times10^{-13}$&P \\
&& 20.0 & 11-50 & 7&--& 0.8 (27) 
&1.6$\times10^{-12}$&5.6$\times10^{-13}$&9.0$\times10^{-13}$&B \\
Source 6&274$\pm19$& 1.68 & 1.56-1.81& 7&-- & 0.8 (45)  
&7.9$\times10^{-13}$&7.4$\times10^{-13}$& 9.2$\times10^{-13}$&P\\
&& 10  & 7-15& 7&--& 0.9 (45)
&8.2$\times10^{-13}$&5.3$\times10^{-13}$& 8.2$\times10^{-12}$&B \\
Source 7&468$\pm23$& 1.87 &1.76-2.00 & 60&42-90  & 1.0 (79)
&2.3$\times10^{-12}$&3.0$\times10^{-12}$&3.3$\times10^{-12}$&P\\
&& 8.0& 7-10   & 40 &25-50 & 0.9 (79)
&2.2$\times10^{-12}$&1.6$\times10^{-12}$& 2.4$\times10^{-12}$&B\\

\hline
\hline
\end{tabular} 

\bigskip
NOTES: Fluxes are from the LECS data only and are calculated for the
best fit parameters given. P stands for Power Law model,
and B for Bremsstrahlung.  N$_H$ is in units of 1\e20 cm$^{-2}$. \\
$^a$  fit to LECS data only gives $\Gamma$=1.20 [0.96-1.44]
and kT=187 [$>$17], for
N$_H$=7$\times 10^{20}$ and $\chi^2_\nu$=1.4
\end{table*}

We have fitted the LECS data jointly with the MECS data.  The addition
of LECS data will not give a significant contribution in the 
overlapping energy region ($>$2 keV), given their lower statistical
weight, but they should add crucial information at low energies.  The
spectral parameters are forced to be the same for all 3 instruments,
while the relative normalizations are free to vary (see Fiore et al. 1999).  
This also takes
into account the fact that a different fraction of the
photons are included in the source area. The low energy
absorption is at first fixed at the Galactic line-of-sight value, and let
free to vary if required by the quality of the fit.  The bulge is 
treated separately (see \S~\ref{thebulge}).

Table~\ref{lecs+mecs} gives the results of the joint fits for the 6
sources detected with the LECS.   In all cases (but source \#~2)
LECS data are consistent with the MECS. For source \#~2,  
LECS data alone  would suggest a higher temperature spectrum (see
Table~\ref{lecs+mecs}) and no intrinsic absorption, while a significant
absorption is suggested in the power law model fit.  This is also the
only source for which the P and B models give significantly different
low energy absorption values.   The absorption parameter measured with
LECS data is consistent with the line-of-sight
values of 7$\times 10^{20}$ cm$^{-2}$ for sources \#~3 and \#~6. 
Sources \#~1 and \#~7 are clearly absorbed: a fit with absorption fixed at the
Galactic value    gives a clear depression at low
energies (and significantly worse $\chi^2_{min}$) 
that disappears with a column density of $\sim 5\times$ higher (see
Fig.~\ref{melespec}).  
This is consistent with the sources being embedded or behind the HI ring
in M31, from which column density of $\ge 3 \times 10^{21}$ cm$^{-2}$ is
expected.  It is also consistent with the results from the $Einstein$
data (TF).

\subsection{The case of the bulge}
\label{thebulge}

Since a single temperature or a single power law model cannot be used
to represent the MECS data for the bulge over the full energy range
(see Table~\ref{spectramecs}), we have tried a different approach.  In
particular, since we expect that the strongback modifies the spectral
distribution of the photons, we have also untied the spectral
parameters, to account for possible differences in the response of the
two instruments.  With a broken power law, we could fit the full energy
range and obtain a minimum $\chi^2_{\nu}$ value of 1.  
To obtain a reasonable value of the
$\chi^2_{min}$, however, the spectral parameters must be significantly
different in the two instruments:  for the MECS2 data we could fit the
full range with a single, steep power law of $\Gamma \sim 2.6$, similar
to what we found for both sets of data at higher energies (or a cut-off
energy of 9.4 keV), but the MECS3 data do require a flatter power law
at lower energies, with $\Gamma \sim 1.4\pm0.1$, and a break (cutoff)
energy at $\sim 5$ keV.   A single power law is never a good fit to the
MECS3 data.  In the assumption of a bremsstrahlung spectrum, we also
can properly fit each set of data, but with very different
temperatures:  $\sim 3$ keV in MECS2 and $\sim 12$ keV in MECS3.  The
two temperature converge to a value around 5-7 keV if only data above 4
keV are considered.

Given the large disparity between the two sets of best fit values, we
cannot interpret this in view of residual faulty calibration between
the two instruments (in agreement to within  a few per cent) 
and therefore we have to interpret this result as 
an indication that there are some more fundamental technical
problems, most likely in the
calibration of the instrument in the vicinity of the strongback and in
the determination of the ARF in cases of such heavy obscuration and
complex morphology (the program assumes a point source distribution of
the photons, for example).  We have tried to understand the origin of
this discrepancy, as briefly explained in ~\ref{app}.  We conclude that 
MECS data cannot be reliably used to derive the
spectral properties of the bulge, except at high ($>$ 4 keV) energies,
where the effect of the strongback is negligible.

\begin{figure}
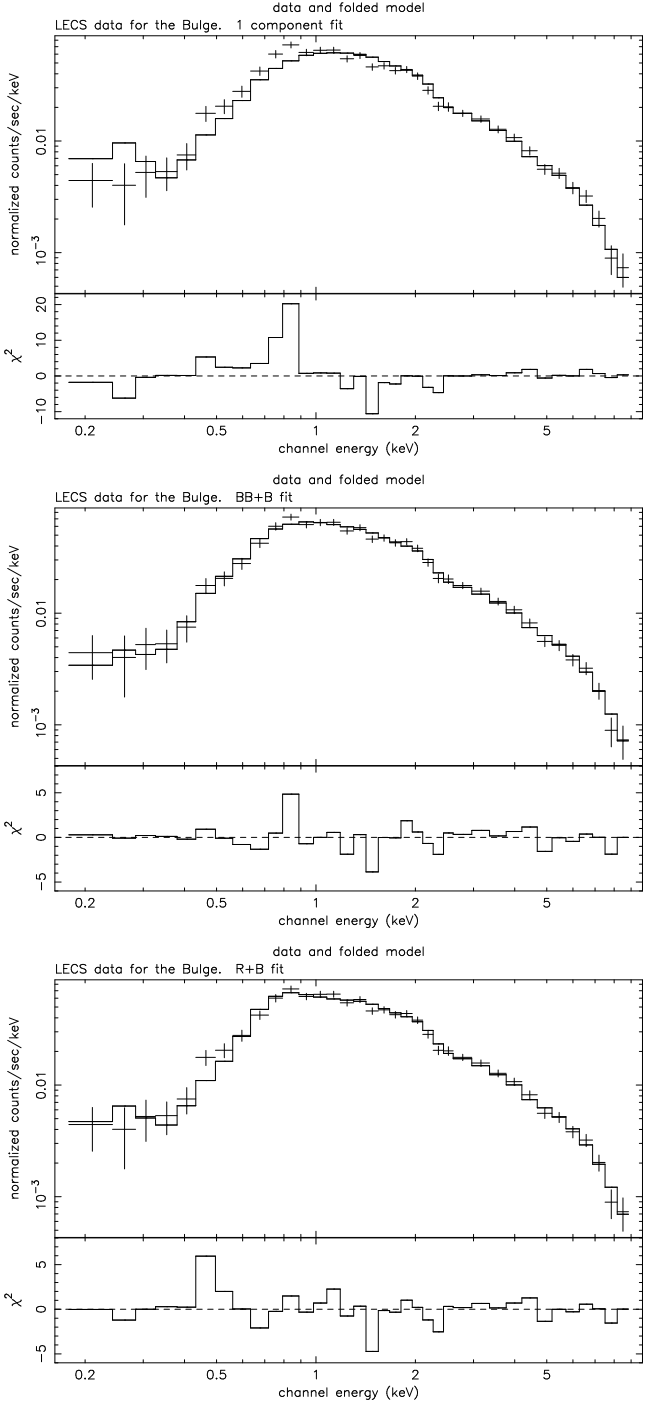

\psfig{file=8757f7a.ps,width=8.5cm,angle=-90,clip=}
\medskip

\psfig{file=8757f7b.ps,width=8.5cm,angle=-90,clip=}
\medskip

\psfig{file=8757f7c.ps,width=8.5cm,angle=-90,clip=}

\caption{Spectral distribution of the photons from the bulge region
detected with LECS.  A single temperature B model is fitted to the data
in the top panel, while a 2 temperature model is used in the middle
and lower panels.}
\label{bulspec}
\end{figure}

In spite of the much shorter observing time, and smaller sensitivity, 
LECS data on the bulge provide high enough statistics to be analyzed
separately from the MECS data, with the added advantage of fewer technical
problems.  The source position, at $\sim 9\farcm5$
off-axis, should make it clear from the strongback in the LECS, 
thus giving us cleaner and independent information on its spectral
properties.  As for other LECS sources, we have built the appropriate
ARF for the area used to extract the source photons
in the point source approximation.  
Table~\ref{bulge} summarizes the relevant results of the spectral
fits to the full spectral range of the
LECS ($\sim$ 0.1-9 keV).    We find that a single power law,
a single temperature bremsstrahlung or a broken power law are
inadequate to fit the data, as shown by the large $\chi^2_\nu$ values, 
since they all leave positive residuals around 0.8 keV
(see Fig. ~\ref{bulspec} top).   To account for this soft excess, 
we have added a component  to the B spectrum.  We have considered a 
Black Body, which, at a temperature of $\sim 0.15$ keV, 
reduces significantly the excess and
the minimum $\chi^2$ value (see Table~\ref{bulge}), although it requires
a higher than line-of-sight value for the low energy absorption.
Fixing the low energy absorption at the line-of-sight value however 
does not change significantly the best fit parameters  (see
Table~\ref{bulge}). 
The residual at $\sim 0.8$ keV involves only one bin, although it
appears significant (Fig.~\ref{bulspec}). 

While this is a good fit to the data, it is not unique.  In fact, a
$raymond$ model (in place of the BB) 
with solar abundances and a best fit kT$\sim$0.3 keV
also reduces both the minimum $\chi^2$ to an acceptable value and the
systematics in the residuals. 
As in the BB+B model, there is a residual positive
excess at $\sim 0.5$ keV that could be significant (Fig.~\ref{bulspec}).

We therefore conclude that the LECS spectra of the bulge can be well
represented with
a two component model,  either BB+B or R+B.  We have further checked 
whether a power law could be used to parameterize the high  
energy component, and found that a B model is preferred 
(see Table~\ref{bulge}),
suggesting a curvature in the photon distribution at high energies.
The addition of MECS data, at energies above 5 keV only, confirms the
results of the LECS data alone.  Given the potential problems related to
the presence of the strongback, MECS data have not been explicitly added to 
the fits of Table~\ref{bulge}.

Although not formally required by the fits, we have nonetheless
attempted more sophisticated model, to further understand the
characteristics of the low energy emission from this region.  We have
released the constrain of solar abundance in the R model.  A better fit
is found for extremely low ($< 1$\%) abundances, but at the expenses of
a very high column density ($\sim 20$\e20 cm$^{-2}$).   A much less
dramatic improvement is found if the column density is fixed at the
Galactic value.  However a
significant decrease in the $\chi^2$ value ($\Delta \chi^2 > 10$ for 1
DoF less) and improved residual distribution is obtained if one of
the elements like N, Ar, or S is allowed to vary, while all others are
kept at the solar value, or if a very narrow line at $\sim 0.5$ keV is
added to the R+B model ($\Delta \chi^2 =12$ for 3 fewer DoF).  In either
case, the F-test probability is $>99.9$\%.  
Alternatively, the addition of a narrow $\sim 0.8$ keV Gaussian line to
the BB+B model has the effect of reducing the requirement of
higher-than-line-of-sight absorption, and improving the $\chi^2$
($\Delta \chi^2 =6$\footnote{The F-test probability is $\sim 99.7$\%. 
Although the improvement is not as dramatic as in
the equivalent case with R+B model, this is due to the lower minimum
$\chi^2$ value in the BB+B model. The final $\chi^2$ value is the same
for both sets of models} for 3 DoF less).  
While these component  might be
physically meaningless, they are nevertheless reminiscent of the more
sophisticated models, such as those used in the data of Her X-1 or
4U1626-67 observed with the LECS (Owens et al. 1997; Oosterbroeck et
al. 1997), that include also line emission at low energies, over the
black-body model, and might be an indication that more sophisticated
models than those of Table~\ref{bulge} should be attempted, when
improved quality spectral data will become available.

\begin{table*}
\caption{Spectral fit results for the bulge source}
\label{bulge}
\begin{tabular}{lrllllllll}
\hline
Model(s)&N$_H$& kT$_s$&90\% err&
\multicolumn{1}{c}{$\Gamma$}&90\% err&\multicolumn{1}{c}{kT$_h$}&90\% err& $\chi^2_\nu$ (DoF)&Notes\\
\hline\hline
\multicolumn{9}{c}{\bsax\ LECS  data} \\
\hline
P&10\e20&--&&1.9&---&&2.0 (33) \\
B&5.8\e20&--&&&&5&---&2.7 (33) \\
BB+B&12\e20&0.13&0.11-0.15&&&6.0&4.7-6.8&1.0 (31)&(1)  \\ 
BB+B&7\e20&0.15&0.13-0.17&&&6.4&5.7-7.3&1.2 (32)& (2) \\
R+B&7.5\e20&0.33&0.25-0.56&&&5.9&5.3-6.6&1.1 (31) \\
BB+P+BB& 10\e20&0.15&0.11-0.18&1.9&1.5-2.2&0.89&0.72-1.12&0.9 (29)&(1)  \\
R+P+BB&11\e20&0.51&0.29-0.70&2.2&1.9-2.5&1.07&0.91-1.20&0.9 (29)&(1)  \\
\hline
\multicolumn{9}{c}{\bsax\ LECS+PDS data }\\
\hline
BB+P+BB& 12\e20&0.15&0.11-0.18&1.9&1.6-2.3&0.92&0.75-1.15&0.9 (35)& (3)\\
BB+P+BB& 11\e20&0.15&0.11-0.18&1.8&1.6-2&0.84&0.75-0.95&0.9 (36) &(4)\\
R+P+BB&10\e20&0.50&0.3-0.7&2.1&1.9-2.5&1.06&0.94-1.2&0.9 (35) &(3)\\
R+P+BB&9.3\e20&0.48&0.3-0.7&1.9&1.8-2.0&0.93&0.82-1.09&1.0 (36)&(4) \\
\hline\hline
\multicolumn{9}{c}{ASCA SIS bright data } \\
\hline
P&2\e20&-&&1.73&1.69-1.77&& &1.2  (276)&(5)  \\
B&$<$1$\times 10^{19}$&--&&&&5.6&5.2-6&1.2 (276) &(5)  \\
BB+B&14\e20&0.11&$<$0.12&&&5&4.8-5.2&1.1 (274) &(6) \\
R+B&7\e20&0.65&0.2-0.8&&&5.7&5-6.6&1.1   (274)\\
\hline\hline

\end{tabular}

\bigskip
NOTES:
Models are: B=Bremsstrahlung; BB = Black Body;
P=Power law; R=Raymond, with fixed
abundances at 100\% cosmic value.  The low energy absorption is 
free to vary in the 0.1\e20 -30\e20 cm$^{-2}$ range. 
kT is in keV; $\Gamma$ is the P photon index. \\
Total counts used in the analysis:\\
5038 $\pm$ 73 counts in the $\sim 0.2-8.5 $keV range (LECS) \\
5640  $\pm$ 898 counts in the $\sim 15-30 $keV range (PDS) \\
32790 $\pm$ 190 counts in the $\sim 0.8- 5$ keV range (SIS) \\
(1) The Galactic N$_H$ value is marginally consistent at the 90\% level \\
(2) The N$_H$ is fixed. \\
(3) The relative normalization between the LECS and the PDS is a free
parameter, and is 2.2 for R+P+BB and 1.4 for BB+P+BB.  \\
(4) The relative normalization between the LECS and the PDS is 
fixed at 1.05 \\
(5) The Galactic N$_H$ value is well outside the range of parameters
allowed by the fit.   Upper boundary of N$_H$  is below 1$\times 10^{20}$
cm$^{-2}$ for B, and 5$\times 10^{20}$ for P. \\
(6) The Galactic N$_H$ value is within the allowed range \\
\end{table*}

\subsubsection{PDS data}

A significant detection in the $\sim 15-30$ keV range 
is obtained in the observation of Field 3 with
the PDS detector. We have used the background-subtracted files provided
by the SAX-SDC, which contains  $\sim 5600$ net counts. 

The large field of view and lack of spatial resolution make it
difficult to identify the PDS source.  The field of M31 is
clearly complex, so there could be one or more candidates from
the MECS sources.
It is also possible that a source unrelated to those detected by the
MECS is responsible for the emission.  However, there is only 1 bright
hard X--ray source in a 1.5$^\circ$ radius around the center of field \#~3 and
it has been associated with M31 since the UHURU days (4U0037+39).  All
other sources detected with imaging missions (for example with the $Einstein$ 
Slew Survey) are significantly fainter.  We therefore suggest that the
PDS detection is due either to a source (a combination of sources) in
M31 or to an unknown, very absorbed, possibly variable background
source.  Since we cannot check on the second hypothesis, we have tried
to further understand whether an association with one or more sources in
M31 is feasible.  

There are two kinds of sources in M31: for the most part they have a
$\sim 5-10$ keV thermal spectrum, but one (possibly two) has a much
harder $\Gamma=1$ power law.  The strongest by far is the source
associated with the bulge region, which can be regarded as a multitude
of sources concentrated in the central part of M31.   The hard source
is significantly fainter than the bulge, but could give a larger
contribution at very high energies.  We expect that if the association
is with a source in M31 it will be with sources in the center or NE
part of the disk. No PDS detection is obtained from the observation of
Field \#~6, which is also in the PDS FoV of Field \#~3.  However, the
upper limit is consistent with a count rate $\sim 1/2$ of that of Field
\#~3,  which could be expected from a source in Field \#~3, that is
detected with reduced intensity due to the lower transmission of the
instrument at large off-axis angles (of the order of $\sim$ 45\% for a
source at the center of Field \#~3)

We have therefore tried a fit of PDS data together with either the 
LECS data for the bulge or the MECS data for source \#~9. 

\begin{figure*}
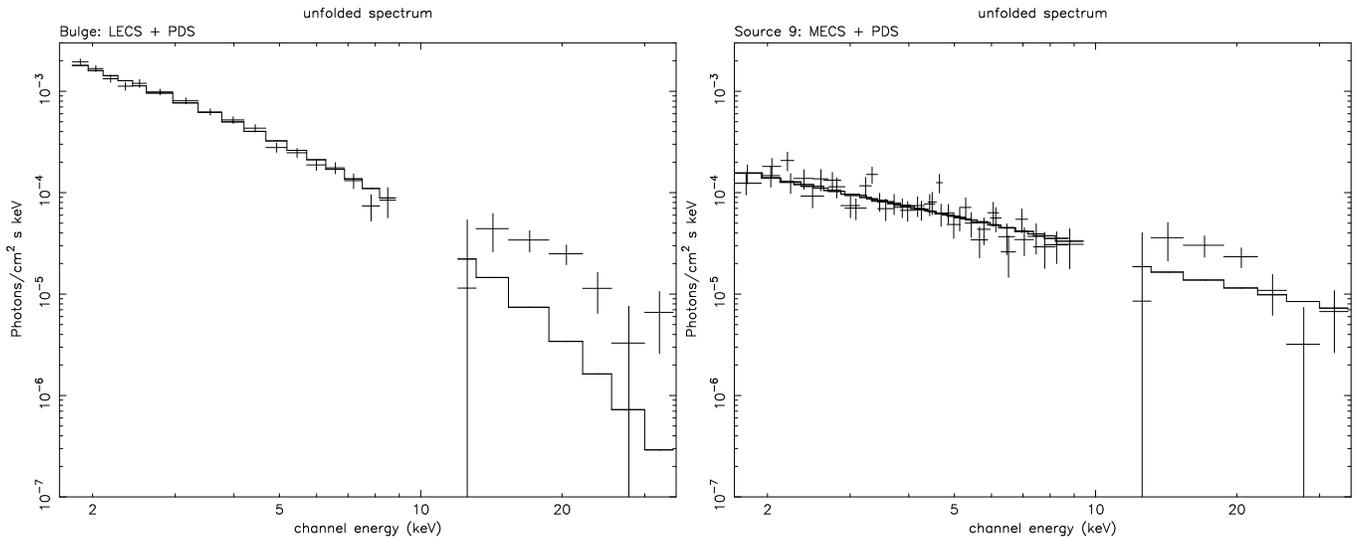

\resizebox{18cm}{!}{
\psfig{file=8757f8a.ps,width=18cm,angle=-90,clip=}
\psfig{file=8757f8b.ps,width=18cm,angle=-90,clip=}
}
\caption{\label{pds}Plot of the  unfolded spectrum and the model
normalized to the MECS or LECS data and extrapolated to the PDS energy
range. LEFT: LECS data for the bulge source, fitted with a $\sim 6$ keV
B model.  RIGHT: MECS data for source \#~9, fitted with a $\Gamma \sim 1$ P
model.  The normalization  for the PDS data is fixed at the maximum
expected value for a source at the field's center (1.2 relative to LECS, 
left; 0.9 relative to MECS,  right;
see text).}
\end{figure*}

We find that if we extrapolate the MECS or LECS results obtained above
to the PDS range, we can account only for a fraction of the detected
PDS counts, as shown in Fig. ~\ref{pds}.  The $\sim$ 6 keV spectrum
that fits the bulge falls $\ge 3.5 \times$ below the PDS detection
(Fig.~\ref{pds}a).  Since the bulge is significantly stronger than the
other sources in the field, the superposition of all their
contributions, if they have the same relatively steep spectrum as the
bulge, will only increase the expectation by less than 30\%, too little
to reconcile the discrepancy (Fig. ~\ref{pds}a).  In Fig.~\ref{pds}b,
the extrapolation of the fit of MECS data for the harder source \#~9
indicates again a factor of $\ge \times 1.4$ discrepancy at the PDS
energy  range.  Both these values are outside the expected
cross-calibration uncertainties (good to $\sim$ 10 \%, Cusumano et al.
in prep.), and much higher than the relative normalization expected
between instruments ($\sim 0.8-0.9$ for MECS-PDS and $\sim 0.8-1.2$ for
LECS-PDS, for sources at the field's center).  
Furthermore, they do not take into account the lower
transmission due to the off-axis position of the sources (that is
reduced at $\sim 0.9 -0.65 $ at 10$'-25'$ respectively).

\begin{figure}
\psfig{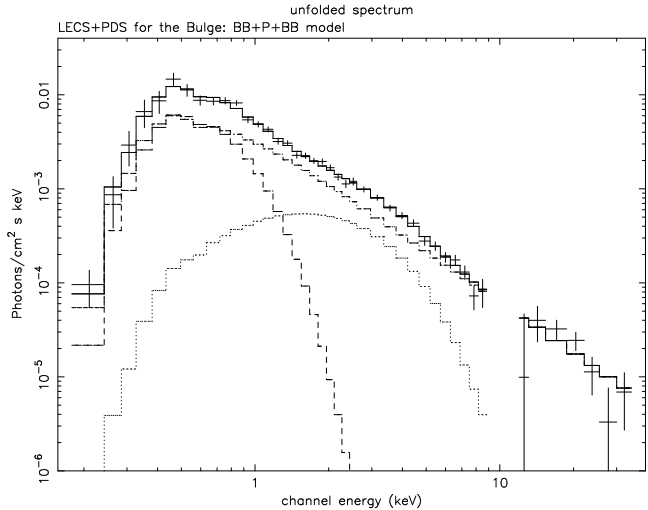}
\caption{\label{allbulge}Plot of the  unfolded spectrum and the model
for the PDS and the LECS data for the bulge region.  The three component
model is a $\sim 0.15$ keV Black Body (dashed curve), a $\Gamma =$ 
1.8 power law (dot-dashed curve) and a $\sim 0.8$ keV Black Body (dotted
curve), with low energy absorption N$_H \sim 11 $ \e20 cm$^{-2}$ (see
Table~\ref{bulge}).  A normalization factor of 1.05 is applied to the PDS
data (see text).
}
\end{figure}

These results would argue against an association with one of the sources
in the MECS FoV, although a combination of sources could account for a
large fraction of it.  However, we notice that if the bulge emission is due to
the contribution of many LMXB, we can use a more appropriate model
than a simple bremsstrahlung to represent the emission at high energies
(White, Stella \& Parmar 1988; Barret \& Vedrenne 1994).  We have
therefore substituted the B with a P+BB model, and fitted the PDS data
together with the LECS data.  We find that with this model the relative
normalization falls to a value of $\sim 1.4$, that, while still higher
than the maximum expected value, is very close to it.
Moreover, as shown by Fig.~\ref{allbulge}, a value of $\sim 1$ (maximum
expected value for a source 10$'$ off-axis) is in very good
agreement with the data. 

As shown in Table~\ref{bulge}, this model is also perfectly adequate
for the LECS data alone.  However, when the P+BB is used for the
high energy data, the BB model for the low energy excess might be
preferable, since the whole BB+P+BB set requires a much lower relative
normalization than the R+P+BB (although, a relative normalization of 1.2
is consistent with the data, see Table~\ref{bulge}). 

Therefore, while we cannot exclude that the PDS detection is the result
of the added contribution of all sources (in particular if they have a
spectrum as hard as the best fit value for source \#9), it can also be
explained as due mostly to the bulge, when the appropriate model for
Galactic LMXB is used to describe the high energy portion  of the
spectrum.

\subsection{Comparison with previous results}

Finally, we have compared the \bsax results with those of previous
instruments.  We expect that the bulge flux and spectrum are constant
in time.  While it is true that each individual source could vary, and
in fact previous analysis on bulge sources have indeed shown
variability (see FT, S97, P93), the spatial resolution of \bsax
prevents us from studying each source individually.  On average
therefore we expect that the global properties of the bulge do not
change (a possible variability in the bulge within this observation is
small, see \S~\ref{variability}), and can therefore be used to
cross-calibrate between different energy bands and different
instruments at different times.

However, the comparison between these and previous results must be done
with caution. 
Imaging instruments like $Einstein$ IPC and ROSAT PSPC
had a much poorer spectral resolution and much narrower energy
band, so we can use them only partially to compare the spectral
properties.  Other missions with good spectral resolution and energy
coverage were non-imaging, so that the results could apply to a larger
area than discussed here. ASCA is the only mission for which we can be
reasonably sure the results apply to the bulge only on a similar energy
range.     Since there are no reports in the literature 
on the observations of M31 with
ASCA, we have obtained the ASCA data from the public NASA archive.     
One observation (sequence 63007000) is pointed almost exactly in
the direction of Field 3, and contains the bulge as well as a few of the
other sources in M31 reported in Table~\ref{bulge}.  We report the
details of the analysis in \ref{asca}, limited to the bulge data. 
The results summarized in Table~\ref{bulge} indicate that 
the ASCA and \bsax\ results are in excellent agreement, both for 
the single B or P model and the  two component fit used for the \bsax\ LECS
results, although the improvement in fit quality is not
as dramatic when a second component is added to the ASCA fits.  This 
partly reflects the more limited extension of the SIS data at low
energies.

We can also compare the present results with previous non-imaging
hard X--ray missions,
that should also be dominated by the bulge emission.  
Fabbiano, Trinchieri and Van Speybroeck (1987) have fitted the $Einstein$ MPC
data ($\sim 2 -10$ keV) with a B model with kT $\sim 6-13$ keV. 
Makishima et al.  (1989) report that  GINGA data instead are not well fit by
simple models: both a cut-off power law or a bremsstrahlung require
high absorbing column.  Therefore  they  suggest a 
model also used to fit the data of the low mass binary population in 
our Galaxy, composed of disk-blackbody and a 
blackbody.  A power law dominating at energies above 10 keV is also
added to account for a possible pulsar contribution. 
As shown in the previous section, simpler models are adequate to
represent the data;  however, the GINGA parameters can also be used,
provided that an additional component is added to account for the
excess emission at low energies.   The results from the bulge colors
derived from ROSAT data also give support to the presence of the soft
component (Irwin \& Sarazin 1998).

The spectral results allow us to determine the flux of the source in
different instruments.  The emitted \bsax\ LECS  flux in the 2-10 keV band
in a 5$'$ radius circle is f$_x = 1.8 \times 10^{-11}$ erg cm$^{-2}$
s$^{-1}$.   This is insensitive to the exact model used (the B or the P+BB
models of Table~\ref{bulge}), indistinguishable in this energy range.  
To compare this with ASCA data we have both estimated the
\bsax flux in the same size region used for SIS, and we have also
obtained the GIS flux in both the larger and smaller regions.  For
consistency, we have applied a 5.6 keV B model to the GIS data as well,
in spite of the fact that this is a poor fit to the GIS data (however,
the 2-10 keV flux does not change significantly if a temperature of 8
keV is assumed).  We find that the total GIS and \bsax flux are in
excellent agreement, while \bsax measures a higher flux than either of
the ASCA instruments in the $\sim 3\farcm2$ radius circle (LECS: $\sim
1.4 \times 10^{-11}$; SIS:  $\sim 6 \times 10^{-12}$; GIS:$\sim 8
\times 10^{-12}$).

To compare it with $Einstein$ and ROSAT values, we have to extrapolate
it to the softer passbands of those instruments.  If we consider the 
single temperature spectrum that fits the data at high energies, we find
f$_x (0.2-4) \sim 2.3 \times 10^{-11}$ and f$_x (0.1-2) \sim 1.6  \times
10^{-11}$ erg cm$^{-2}$ s$^{-1}$ with \bsax, slightly smaller than reported 
by FT and S97, who use equivalent spectral models.  
On the other hand, the spectral fits indicate that a single
temperature model fails to represent the data at low energies, and we
find higher fluxes when we consider more complex models.  

Non-imaging instruments give also a somewhat higher  flux.  Makishima
et al. report a total flux of $\sim 8\times 10^{-11}$ erg cm$^{-2}$
s$^{-1}$ in the 2-20 keV band from GINGA.  However, if all of the
emission measured by GINGA is due to M31 only, and the rough factor of
2 between bulge and total luminosity observed at softer energies
holds also at higher energies, this would imply a flux of $\sim 3
\times 10^{-11}$ erg cm$^{-2}$ s$^{-1}$  in the 2-10 keV passband.

Given all of the uncertainties involved, we can probably safely assume
consistency between all of these values.  This ensures us that we can
estimate the flux of the bulge, which we assume to be constant, in the
\bsax data, and use it to better evaluate the quality of the measured
flux in other sources that could suffer from similar problems, to
compare them with fluxes obtained with other missions and study flux
variations at different epochs.

\begin{table}
\caption{Results from the timing analysis of MECS data}
\label{var}
\begin{tabular}{lccccc}
\hline
\hline
Name & Cts  &Long--term &  \multicolumn{3}{c}{Upper limits $^a$
} \\
     & used&  Variability  &  10$^4$\,s & 10$^3$--10$^2$\,s &
10--5\,s \\
\cline{4-6} 
     & (\#)& ($\chi^2_{\nu}$$^b$) & \multicolumn{3}{c}{(\%)} \\       
\hline
\hline
Source 1      & 1276& 1.15 & 35 & 36--38 & 37--38  \\
Source 2      & 3150& 1.30 & 24 & 24--23 & 23--25  \\
Source 3      & 1288& 1.13 & 38 & 35--36 & 40--43  \\
Source 4      & 1049& 0.85 & 38 & 40--42 & 45--49  \\
Source 5      & 8523& 2.45 & 15 & 15--14 & 16--17  \\
Source 6      & 1431& 1.05 & 33 & 34--33 & 38--42  \\
Source 7      & 3256& 1.47 & 27 & 23--22 & 25--28  \\
Source 8      & 1481& 1.21 & 32 & 34--33 & 37--40  \\
Source 9      & 1071& 1.02 & 43 & 38--39 & 43--50  \\
Source 10     & 825 & 0.95 & 50 & 44--45 & 50--60  \\
Source 11     & 854 & 1.17 & 44 & 44--43 & 50--55  \\
\hline
\hline
\end{tabular}\\

Notes:
\begin{description}
\item{$^a$} at the 99\% confidence level
\item{$^b$} DoF = 80 for sources 1 and 2, 170 for sources 3 to 11.
\end{description}

\end{table}

\subsection{Source variability}
\label{variability}

Among the most luminous persistent X--ray sources in our
Galaxy are the LMXBs. These sources often
show a large flux variability on long  timescales (from days up to
years) and are characterized by relatively short orbital periods (of
the order of hours), the modulation of which is also detected at X--ray
energies (see White, Nagase and Parmar 1995).
Similar objects are expected to be
seen in M31 and a search for periodic and aperiodic variability was
therefore carried out. 

We extracted the photon arrival times for each source from a circular
region corresponding to the 90\% of the encircled energy of the merged
data of MECS2 and MECS3.  We performed a search for both periodic and
aperiodic variability in the following way.  We first accumulated
1000\,s binned light curves for each source and searched for variations
such as increases, decreases or impulsive variations within the time
interval covered by the observation, through the comparison with a
constant. All the sources but one are consistent with being constant
(see Table ~\ref{var} for details).  Source \#~5 is the only one
showing a relatively high $\chi^2_{\nu}$.
However, this corresponds to a $<$10\% flux variation, probably close to
5\%,  suggesting that at most 1-2 of the $\sim$ 50 bright sources
detected in the high resolution images have varied within the
observation.   Caution should also be used in interpreting this flux
variation, since the close proximity to the strongback might
introduce some unknown low level effects related to the small scale
motions of the satellite, although there is no evidence of
a satellite drift during this observations.  

After converting the arrival times to the Solar System barycenter,
we searched for a sinusoidal modulation in the X--ray flux of the
sources.  We have accumulated light curves binned in 0.5\,s and
calculated a single power spectrum for each source over the whole
observation.  We adopt a recently developed technique (Israel \& 
Stella 1996) aimed at the detection of coherent and quasi--coherent 
signals in the presence of additional non--Poissonian  noise 
component in the power spectrum, while preserving the Fourier 
frequency resolution. In this technique, the continuum
components of the spectrum at the j--th frequency are evaluated based
on a logarithmic smoothing which involves averaging the spectral 
estimates adjacent to the j--th frequency over a given logarithmic interval 
excluding the j-th frequency itself. 
By dividing the sample spectrum by the smoothed one a white--noise like 
spectrum is obtained, the approximate probability distribution function 
of which is derived based on the characteristics of the sample 
spectrum. A search for coherent pulsations is then carried out by
looking 
for peaks in the divided spectrum, for which the probability of chance 
occurrence is below a given detection level. If no significant peaks are 
found, an upper limit to the amplitude of a sinusoidal modulation is 
worked out for each searched frequency. 

No significant periodicity was found in any of the power spectra above
the 99\% confidence threshold.  In Table~\ref{var} the corresponding
upper limits to the pulsed fraction for selected trial periods are
shown.

\begin{table}
\caption{Comparison of mean fluxes in \bsax, $Einstein$ and ROSAT}
\label{flux}
\begin{tabular}{lcccrc}
\hline
\hline
&\multicolumn{4}{c}{2-10 keV flux } \\
Name&\bsax&$Eins.$&\multicolumn{2}{c}{ROSAT} &Notes\\
&&&PSPC&HRI \\
&\multicolumn{4}{c}{($\times 10^{12}$ erg cm$^{-2}$ s$^{-1}$)}\\
\hline
\hline
Source 1& 4.6& 2.5 &1.8   & \\
Source 2& 2.2& 1.6  &1.3  &  \\
Source 3& 1.3 & 0.9  &1.2 &&1\\
Source 4& 2.5 & 0.4 & 1.0&0.6&2,3 \\
Source 6& 1.2 & 0.3 &0.7&0.5&2 \\
Source 7& 3.6 &1.2 &2.3 &1.5&2\\
Source 8& 1.6 &1.2 & 1.7 &1.8&1\\
Source 9& 3.4 & 0.4   & 0.3  &0.4&4  \\
Source 10& 1.3 &1.0 &1.0  &1.0&1,2\\
Source 11& 1.2 &0.9 & 0.8  & &1\\
\hline
\hline

\end{tabular} 

\medskip
Notes:
\begin{enumerate}
\item Near the strongback and/or at large off-axis (see text). 
\item Given the possible association with more than 1 ROSAT ($Einstein$)
source, the flux of all is reported for comparison with the \bsax\ flux. 
\item Source at the edge of the field
\item Flux of Source \#~9 is in 4$'$ radius circle (see text). 
\end{enumerate}

\end{table}

To search for long term variability we compared the flux measured at
different epochs by different instruments.  Table~\ref{flux} shows the
comparison between MECS fluxes and the average fluxes obtained with
$Einstein$ and ROSAT for the different sources detected
with \bsax.  We have converted the 0.2-4 keV fluxes given in FT, 
S97 and P93 to a 2-10 keV flux assuming a $\sim 6$  keV Bremsstrahlung model.
When more than 1 ROSAT ($Einstein$) source is included in the count
extraction region, the sum of all fluxes is given 
in Table~\ref{flux}.  The adopted
count-to-flux conversion of Table~\ref{spectramecs} are expected to be
reasonably accurate.
However, the flux of sources at large off-axis angles or
near the strongback could be under/overestimated, since the ARF (which
properly models the expected spectral distribution of the photons, as
already discussed) does not take into account distortions at large
off-axis angles and does not properly correct for the strongback
absorption.  This could lead to an overestimate of the flux for sources
near the strongback (although probably $\le$ 40\% in the worst case, and
our sources are only partially affected by the strongback),
and to an underestimate for very off-axis sources, in particular as a
result of the small area that we had to use due to field crowdedness.
In fact, the flux for source \#~9 derived from a 4$'$ radius circle is
higher by $\le 50$\% than that obtained from the $2\farcm6$
circle reported in Table~\ref{spectramecs}.
Unfortunately, this is the only source at large off-axis angles for
which a larger area can be used to test this.  All other sources
(namely \#~8, \#~10 and \#~11) are either close to the strongback or to
other \bsax\ sources.

The comparison in Table~\ref{flux} indicate that \bsax\ fluxes are
systematically slightly higher than either ROSAT or $Einstein$ fluxes.
However, there appears to be  a roughly constant factor of $\sim \times
1.5$ between \bsax\ and $Einstein$ fluxes, regardless of source
position in the field, which would point to a further systematic
off-set, rather than a flux increase for all sources.    In fact, if we
consider that most of the sources are close to or embedded in the HI
disk, and that absorption effects are much more important in the softer
energy bands of $Einstein$ and even more of ROSAT, it is likely that
neglect of the internal absorption in M31 in the counts-to-flux
conversion in the softer enegy bands (both FT and S97 have assumed only
absorption equivalent to the Galactic line-of-sight value) accounts for
most of this off-set.

Three  sources however deviate from this trend: source \#~9 is
much stronger in the \bsax\ data of Dec. '97 than in previous
observations, and sources \#~6 and (less drammatically) \#~7 
are stronger than measured by $Einstein$ (ROSAT fluxes are consistent
with a increase since then).     Their location in M31 indicates
that absorption could be severe if they are in or behind the HI ring.
However, this would not be sufficient to reconcile the different
fluxes.  Moreover, only the spectrum of source 9 is significantly
different from the one adopted in the flux-to-counts conversion, and
again this is not  enough to bring $Einstein$ or ROSAT
fluxes to the \bsax level.  It is therefore likely that these sources
have varied in the $\sim 20$ years elapsed between observations.  A
better assessment of the amplitude variation will however require a
more precise knowledge of the spectrum, which will be possible with
future, broad band observations such as those available with the AXAF
or XMM missions.

\section{The Globular Cluster sources in M31}

All M31 source detected with \bsax\ 
have high X--ray luminosity (L$_X \ge 5 \times 10^{37}$
erg s$^{-1}$ in the 2-10 keV band), and have been identified mostly
with globular clusters.  This suggests that they are  most likely Low
Mass X-ray Binary sources.  Although the quality of the data does not
allow us a precise assessment of their spectral properties, we find
that most of the high luminosity sources have a similar spectrum, that
can be described with a single temperature component with kT$\sim 6-9$
keV.  Two sources however have significantly different spectral
properties:  source \#~3 and \#~9, both identified with globular
clusters, have a much harder spectrum, with $\Gamma \sim 1-1.4$.

Although detailed observations of high signal-to-noise Galactic sources
might require more complex models, the spectrum of a LMXB, with a
weak-field neutron star as the accreting object,  is reasonably well
approximated by a Bremsstrahlung model from a few to $\sim 100 $ keV
(see van Paradijs 1998 and references therein).  
In globular cluster sources, where LMXB are
expected,  a range in temperatures, from $\sim 6-20$ keV has been found
from archival EXOSAT data (Callanan et al. 1995).   This is the same
range of temperatures we find for the globular cluster system in M31,
with the possible exception of one source (\#~9).  Therefore it appears
that the spectral properties of the globular clusters in M31 and in our
Galaxy in the $\sim 2-10$ keV band do not differ significantly.  
To better model the low energy data, that cannot be reproduced simply by
the effect of absorption, Callanan et al. also include a BlackBody component 
with kT $\sim 0.5-1$ keV.  As shown by
Table~\ref{lecs+mecs}, \bsax data do not require additional 
components, since a single P or B model plus absorption is adequate in most
cases.  The addition of a BlackBody component would in some cases reduce
the requirement of high absorption, but without improving the quality of
the spectral fit and without reconciling the N$_H$ to the
line-of-sight value (for example, the absorption for source \#~7
is reduced to 28\e20
cm$^{-2}$, if a $\sim 1$ keV BB is added to the P model, see
Table~\ref{lecs+mecs}).   

The sample examined by Callanan et al. spans a rather large range in 
X--ray luminosities (from 5$\times 10^{35}$ to 5$\times 10^{37}$ erg
s$^{-1}$), while the globular cluster sources in M31 are all bright
sources (L$_X \ge 5\times 10^{37}$ erg s$^{-1}$). 
All of the sources studied by Callanan et al. have metallicities lower
than 1/2 solar,  while the \bsax globular clusters have metallicities up
to $\sim \rm solar$ (Huchra et al. 1991). 
It has been recently proposed by Irwin \& Bregman (1999) that the soft
X--ray properties of the globular cluster systems in M31 depend on
metallicity, in the sense that the spectra become softer with increasing
metallicity.  No such trend was found in the Galactic globular clusters,
however Irwin \& Bregman suggest this is due to the lower average 
metallicity considered.   Like for the Galactic clusters, no trend 
is observed between the 2-10 keV spectra of our sources and metallicity:  
the same best fit temperature is
derived for clusters at the opposite end of the metallicity range. 
Although the sample is limited (more so than the ROSAT sample studied by
Irwin \& Bregman) and spans a somewhat narrower range in metallicity
(they have 1 object with higher metallicity), we cannot extend their
suggestion to the harder energies.   We have also considered
the softer energy band,  where however the sample is further reduced both 
in numbers (3 objects) and in metallicity (all metal poor).  As
discussed above, the \bsax data do not require a second
component in the fit.   While this is probably due to the data quality,
it could again be interpreted in the
framework of metallicity:  we have LECS data only for the lower
metallicity objects, and if the requirement of a second component
is not as stringent for these objects, our 1-component fits are
consistent with the low metallicity globular cluster population 
of our Galaxy.  

We have also compared the best fit spectral parameters derived
from ROSAT and \bsax data.  The comparison is not straightforward, 
given the almost completely separate wavebands considered,
also in view of the supposedly complex spectrum of these sources.  
Nontheless, we find that the results are in good, though loose,
agreement.  
The higher than Galactic absorption required by the fit of sources 2 and
7 is also detected in the ROSAT data (ROSAT source 73 and 205
respectively have the highest values of N$_H$ in the Irwin \& Bregman
sample).   There is a much looser agreement with the temperatures;
however, the determination of temperatures such as those measured in
these sources is very hard with ROSAT data.  We notice however that
the spectra of Irwin \& Bregman can be divided in two classes: hard (kT
$>$ 3 keV) and soft (kT $\sim 1-1.5$).  While we do not have any
evidence for the soft spectra, it is possible that they represent the
soft component that we do not measure in our data, either for lack of
LECS data (source \#~8) or possibly because of confusion in the presence
of high absorption (source \#~2).  
Given the extremely limited size of the sample, and the limited quality
of our data, we have to wait for future observations of M31 to really
better measure the spectral properties of its globular cluster
population in the entire $\sim 0.1 -10 $ keV band. 

Source 9 has a much harder spectrum than all other sources in M31,
and in particular it is harder than all other globular cluster sources.    
Hard spectra such as these are more typical of binary systems containing
a strong-field neutron star, or black hole candidates.    
This is a rather unusual spectrum for a globular cluster source, 
as none are known in our own Galaxy.   
We therefore suggest two possible interpretations:  either the surce
has been incorrectly associated with a globular cluster, or this is the
first evidence of black hole formation in a globular cluster.  While
this latter would be a more intriguing possibility, we cannot at the
present time rule out a mis-identification.  A precise determination of
the X-ray position of this very hard source will be possible with
future imaging telescopes and will allow us to confirm its
identification with the optical counterpart.

\section{The bulge of M31}

We have measured the spectrum of the M31 bulge as a whole.  We find
that a single temperature thermal model can represent well the LECS data at
high energies up to $\sim 9$ keV, but fails to account for excess
emission at low energies.  Moreover, if the detection at $\sim 15-30 $
keV obtained with the PDS is associated with the bulge, 
a more complex model is needed also at high energies.  
Unfortunately, the data quality does not allow us to uniquely identify
the different components required to fit the entire $\sim 0.2 -30 $ keV
range of data.

\noindent{\sl High energy emission}:  Until imaging data at high
energies are available, the association between the emission at $\sim
15-30$ keV and the bulge cannot be confirmed.   When a
combination of power law and Blackbody (also used for the Galactic
LMXB) is fitted to the LECS data of the bulge, the PDS data appear as
the extension at higher energies of the bulge emission, indicating that
the association is at least likely.  However, we cannot unambiguously
determine the spectral models to describe the data:  a power law plus
black body  is sufficient to model the LECS + PDS data, but 
the BB+DiskBB+P model, with the parameters used for GINGA
data,  could also describe \bsax data.

\noindent{\sl Soft excess}: 
Within the LECS data, we can model equally well the 
soft component with either a $\sim 0.15$ keV  Black Body or a
$\sim 0.3$ keV  $Raymond$ model.  However, when the Power Law + Black
Body is used at high energies, the Black Body might be slightly preferred.  

We have tried to understand whether the model used for the bulge is
consistent with that of Galactic sources.   The much better quality
spectra that can be obtained for these latter complicates the
comparison, since more detailed complex models are needed to fit the
data.  On the other hand, the much higher typical line-of-sight column
densities of LMXB in the disk of our Galaxy (with the exception of those 
nearby) prevent a proper study of their soft spectra.  
Two of the most nearby  Galactic LMXB's (Hercules X-1 and
4U 1626-67) have been recently observed
both with ASCA and with \bsax.   Their spectrum
needs at least two components; in either case  a Black
Body at low energies and a power law have been used. 
Residual excess emission around   
$\sim $1 keV has been modeled with Fe lines in
Hercules X-1 (Oosterbroeck et al. 1997) and with O and Ne lines in 
4U 1626-67 (Owens et al. 1997).  
The LMXB in globular clusters also require a 
two component model, composed of a Black Body and Bremsstrahlung
or power law component (Callanan et al. 1995).  

The requirement of two Black Bodies in the M31 data is due to the need of
accounting for both excess at low energies and for the high energy
emission, while retaining enough curvature in the spectral shape to  
be consistent with the energy distribution of the photons.  

The temperature of the  softer Black Body component ($\sim$ 0.15
keV, see Table~\ref{bulge}) is intermediate between Her 
X-1 and 4U 1626-67 (kT $\sim$ 0.1 and
0.3 keV respectively, Oosterbroeck et al.  1997; Owens et al. 1997),
and represents a similar percentage of the total 0.1 - 10 keV flux.
However, it gives a much smaller contribution if softer energy bands are
considered ($i.e.$, in the ROSAT band, the unabsorbed flux due to the
Black Body is 30\% of the total flux, compared to $\sim$ 50\% 
in Her X-1).  The hard part of the spectrum is however significantly
different, in particular it is much softer than in the two LMXB, and
more reminescent of the spectra of the globular cluster sources. 
Line emission, that has been recently added to the spectra of disk
LMXB, is not formally required by our data.  However, this 
might only be a limit of the data quality, rather than an intrinsic  
difference between the two groups of sources.  

It therefore appears that the spectral properties of the bulge reflect
both the disk and the globular cluster LMXB properties (assuming that Her X-1
and 4U 1626-67 are typical of disk LMXB, which they could not be, since
they are pulsating sources).  This result is not surprising, since several
sources contribute to the bulge emission, and a mixture of disk and
globular cluster LMXB is to be expected, given the proposed
identifications (TF; S97; P93).

From the normalizations of the Black Body models, we derive a similar
luminosity L$_x \sim$ 4$\times 10^{38}$ erg s$^{-1}$ in both
components, and surface areas r$^2$ $\sim 8 \times 10^6$ and $\sim
5000$ km$^2$ for the soft and the hard components respectively.   The
parameters for this latter are quite reasonable, and suggest the
presence of $\sim 50$ neutron stars in the area, consistent with the
imaging data (TF, P93).  The parameters of the soft Black Body are less
clearly understood.  The luminosity would suggest the presence of $\sim
1000$ Her X-1 type sources (assuming a Black Body luminosity of
6$\times 10^{35}$ erg s$^{-1}$, Dal Fiume et al. 1998), each with a
radius of $\sim 100$ km (which is larger than the radius of the
neutron star in the system as this component is thought to be due to
reprocessing in the accretion disk).  This is in contrast both with the
number of neutron stars derived from the hard data, with the total
luminosity and with the shape of the hard spectrum.  However, until we
can precisely assess the proper model for the soft component, we cannot
reliably determine its intrinsic parameters.

Alternatively, we could consider whether  the soft excess could be
attributed to the diffuse emission apparent in the ROSAT bulge image,
that P93 do not attribute to individual lower luminosity sources.  
P93 estimate that $\sim 30$\% of the total bulge
luminosity could be attributed to either a new class of sources or to a
hot interstellar medium.   In this latter case, it would most likely have 
a plasma spectrum.  
In our analysis however we find that the $raymond$ component contributes
$\sim$ 15\% of the bulge luminosity in the ROSAT band, and would
therefore only account for 1/2 of the residual emission.  Furthermore, 
this interpretation poses
limits to the presence of Her X-1 type sources from the bulge, since
they also appear to contribute significantly to the soft band.   
Spatially resolved spectra of the bulge are needed to clarify the issue
further.

Irwin \& Sarazin have recently suggested that LMXB sources could be
entirely responsible for the soft X--ray emission detected in the
X--ray faintest early type galaxies.  They suggest that the colors of LMXB
and of the bulge of M31 determined within the ROSAT band are in
excellent agreement with those of the low L$\rm _X/L_B$ objects, and
that consequently the  need to resort to additional components (stellar
coronae, a hot interstellar medium) are significantly reduced.
While the presence of at least 2 components in the bulge data has been
established, with roughly the correct parameter values, 
which would support Irwin \& Sarazin's proposal, 
the relative contributions appear to be different from what is measured
in low X--ray luminosity early type galaxies.   

In early type galaxies, the soft and hard components contribute almost
equal amounts in the 0.1-2 keV (ROSAT) band.  In the harder $Einstein$
band (0.2-4 keV) the hard-to-soft ratio is $\sim 2$ and becomes $\sim
4$ in broader, harder bands (Kim et al. 1996; Fabbiano, Kim \&
Trinchieri 1994). In the R+B model (used for the early type galaxy
spectra), the hard component contributes $\sim 5\times,~10\times,~10
\times$ the soft component in the three bands respectively  (although
similar ratios are found also in the ASCA results, this cannot be used
as a strong support, since ASCA data do not formally require the soft
component).  This would suggest that while qualitatively similar, the
spectrum of the M31 bulge cannot entirely reproduce the spectra of low
X--ray luminosity early type galaxies, that require an additional
component over the pure LMXB contribution.  The positive detection of
gas in one of the low L$\rm _X/L_B$ early type galaxies, NGC 1316 (Kim
et al. 1998), further reinforces the need of more than just binaries in
these objects.  Clearly, the presence of a soft component in the
spectral properties of LMXB will have to be properly taken into account
to correctly measure the contribution from an additional soft component
in early type galaxies.  On the other hand, the present observation
shows that the spectral analysis of sources as complex as the bulge of
M31, in which the contribution of several different components and/or
objects are expected, requires high signal to noise data over a large
energy range, to properly assess the individual contributions, and
correctly interprete the origin of each of them.  It is to be expected
that the forthcoming high throughput and high spatial resolution
missions such as XMM and AXAF will give us the wealth of data necessary
to properly address the study of complex sources such as galaxies.

\section{Conclusions}

We have measured the spectral characteristics of 10 individual 
sources and of the bulge region in M31.  

Most of the sources we have detected are identified with globular
cluster, and they appear to have spectral properties consistent with
those of the Milky Way sources.  One of them however appears to have a
significantly harder spectrum, uncharacteristic of LMXB with a weak
field neutron star as the accreting object.  Since High Mass X--ray
binaries are extremely unlikely in globular clusters, we propose that
either this is a mis-identification, or that the LMXB is a black-hole
candidate.  This would be the first such object detected in globular
clusters.

The bulge of M31 as a whole has a multicomponent spectrum.  At high
energies, it is well modeled with a LMXB spectrum, consistent with the
high resolution images that suggest the dominant presence of many
individual sources in the area.   At low energies,  however, an
additional component is needed to model excess emission below $\sim 1$
keV, also possibly associated with the LMXB disk population of the bulge.

High energy emission is detected at $\sim 15-30$ keV with the PDS
instrument.  It is likely that a major fraction of this emission results 
from the M31 bulge, although a contribution from other M31 sources can
also be expected.

\appendix

\section{MECS data for the bulge}
\label{app}

As already discussed,  due to the configuration of the \bsax\ satellite, 
the positions of the sources are different in
the two MECS instruments.  In particular,  the peak of the bulge emission
is located right under the circular structure of the
strongback in MECS2, while it is at a smaller off-axis angle in MECS3.  The
correction applied to the two sets of data are therefore different. 
On the other hand, the customized ARF that we have produced takes into account
the effects of the obscuration from the ``strongback", as shown by the 
shape of the spectral models folded through the instrument response. 
Moreover, the reliability of the ARF has been further confirmed also on
the spectral analysis of a couple of  pulsars that are located at
different off-axis angles in different observations, as already
discussed earlier.

The corrections included in the ARF assume that the photons are
distributed as a point source and refer to the peak position.  This
would suggest that the effective area file produced for MECS2
simulates more accurately the effects of the obscuration from the
strongback than  that produced for MECS3, for which a more
sophisticated model for the spatial distribution of the photon should
be used, to take into account the fact that a larger fraction of
photons than expected based on a PSF model are obscured at off-peak
positions.  We also noticed that, if this assumption is correct, the
flattening observed at lower energies in MECS3 is consistent with a
heavier absorption, not properly corrected.

On the other hand, data obtained in a region covering the bulge
emission but small  enough to be ``free" from the strongback
contamination, should provide a cleaner way of determining the spectral
parameters of the bulge (if we choose to remain inside the strongback,
where calibrations are better, we are in a situation analogous to
source \#~7 w.r.t. the strongback).  In MECS3 we can define a circle
centered on the peak position of the bulge emission, while in MECS2 we
can define only regions at the outskirts of the bulge.  However,
if the spectral parameters are uniform across the entire region this
should not introduce additional parameters.  We find that the spectral
results obtained from these smaller areas are consistent with those
from the whole source in the same instrument.  This suggests that there
are residual calibration problems even in regions ``free" of the
strongback (caution however should be taken in defining a region as
``free" of the strongback, since the boundaries of its
effects are not sharp, and obscuration in its
vicinities also also depend on the stability of the satellite during
the observation).  As already mentioned, the effects of the strongback
on the spectral distribution of the photons from a point source are
accounted for.   We conclude that the extended and complex
morphology of the source is responsible for the failure to reconcile
the spectra from the two MECS, since it is likely that a very accurate
and specific modeling not available at the present time is required to
reproduce  the effects of obscuration and scattering produced by the
strongback.   

We have further checked the above considerations with LECS data.  As
discusses in   \S~\ref{thebulge}, LECS should provide a cleaner
set of data for the bulge region. 
For a direct comparison with the MECS data, we have used the 
LECS data in the $\sim 2-9$ keV range. 
A  single power law model gives $\Gamma \sim 2$
(intermediate between the two MECS), but it is not a good fit
(minimum $\chi^2_\nu$ =1.5 for 14 Degrees of Freedom (DoF)). 
A broken power law  or a bremsstrahlung model significantly improve the
minimum $\chi^2$ value ($\Delta \chi^2 > 11$), again with 
best fit parameters intermediate between the single MECS values: 
$\Gamma_1 \sim 1.4$,  $\Gamma_2 \sim 2.4$, E$_B \sim 3.5$,  $\chi^2_\nu$
= 0.8 for 13 DoF; or kT$\sim
6$ keV,  $\chi^2_\nu$ = 0.8 for 14 DoF.

\section{ASCA data}
\label{asca}

We retrieved the screened data files processed with REV2 from the
archive.   We have used data from all 4 instruments without further
cleaning of the data, and selected the ``bright" data mode for SIS0 and
SIS1. This results in $\sim 89$ ks exposure for GIS, and $\sim$ 76 ks
and $\sim$ 81 ks for SIS0 and SIS1 respectively.
Spectral data have been extracted in circles, centered at the
peak position of the X--ray source coincident with the bulge.  For GIS data, 
we have selected a circle of 20 pixel radius ($\sim 5'$), 
while 
for SIS data we have used a smaller circle of 30 pixels ($\sim
3\farcm2$) so that the source region is entirely contained in the CCD
chip.  This causes a problem in the flux determination, but
should not affect the spectra if the characteristics are the same
throughout the region.  
The background was obtained from a
circle of the same dimension at the same detector position from the
blank sky fields also available from the ASCA archive.  
ARF files have been obtained with the $ascaarf$ routine in $ftools$ and
the appropriate RMF have been obtained from the archive for GIS and
build with $sisrmg$ for SIS. 

We have used the $\sim 0.9-9$ keV range for GIS and $\sim 0.8-5$ keV for
SIS, to restrict ourselves to the best calibrated energies.
We have then followed the same procedure as for the \bsax data, first on
the SIS and GIS separately. We have forced the spectral parameters to be
the same in different instruments
but let the normalization free.  At high energies, the data can be fit
by a thermal bremsstrahlung model, but the best fit temperatures  of GIS and SIS
are significantly different, higher for GIS than for SIS.   An excess
over a single temperature model is present at low energies.  A 
discrepancy between GIS and SIS was already noticed in the spectral data
of 3C 273, reported in the comparison of ASCA/XTE/\bsax results (Yaqoob
et al. 1997), although the discrepancy goes in the opposite sense than
here.   We checked that this result is not due to the different
extraction regions (smaller for SIS) by extracting the GIS in the same
size circle, and found almost identical best fit values.  Since ASCA-SIS, 
\bsax-MECS and XTE-PCA agree in the case of 3C~273, we have decided to
use only SIS data in this comparison.  The results of the spectral fits
are reported in Table~\ref{bulge}.

\begin{acknowledgements}
This work has received partial financial support from the Italian 
Space Agency.
GT thanks Prof. Tr\"umper and the MPE for hospitality while part of this
work was done.   TB was supported by NWO Spinoza grant 08-0 to E.P.J.van
den Heuvel.  The SAX-SDC team has been extremely useful, 
cooperative and kind in leading us through the complexity
of the \bsax instruments.   We thank the referee, Dr. J. Irwin, for
useful comments that led to an improved version of the paper. 

\end{acknowledgements}

%
%

\end{document}